\documentclass[fleqn,10pt]{wlscirep}
\usepackage[utf8]{inputenc}
\usepackage[T1]{fontenc}

\usepackage[english]{babel}

\usepackage{amsmath}
\usepackage{amssymb}
\usepackage{amsfonts}
\usepackage{mathtools}
\usepackage[binary-units=true]{siunitx}
\usepackage{wasysym} 
\usepackage{caption}
\usepackage{subcaption}
\usepackage{graphicx}
\usepackage{xcolor, colortbl}

\usepackage{booktabs}
\usepackage{tabularx}

\usepackage{tikz}
\usetikzlibrary{external,automata,trees,shadows,matrix,fit,decorations,positioning,shapes,plotmarks,patterns,arrows,calc,quotes,tikzmark,arrows.meta}

\newcommand{\mat}[1]{\ensuremath{{\mathbf{\MakeUppercase{#1}}}}}

\makeatletter
\renewcommand{\vec}[1]{%
	\ifcat\relax\noexpand#1%
	\ensuremath{\boldsymbol{\lowercase{#1}}}%
	\else
	\ensuremath{\mathbf{\lowercase{#1}}}%
	\fi
}
\makeatother

\makeatother

\makeatletter

\newcommand{\transpose}[1]{\ensuremath{{#1}^{\textsc{t}}}}

\newcommand{\inverse}[1]{\ensuremath{{#1}^{-1}}}

\newcommand{\R}{\ensuremath{\mathbb{R}}}

\newcommand{\norm}[1]{\left|\left|#1\right|\right|}

\usepackage{pifont}

\title{A Direct Comparison of Simultaneously Recorded Scalp, Around-Ear and In-Ear EEG for Neural Selective Auditory Attention Decoding to Speech}

\author[1,2,*]{Simon Geirnaert}
\author[3]{Simon L. Kappel}
\author[3]{Preben Kidmose}
\affil[1]{KU Leuven, Department of Electrical Engineering (ESAT), STADIUS Center for Dynamical Systems, Signal Processing and Data Analytics, 3001 Leuven, Belgium}
\affil[2]{KU Leuven, Department of Neurosciences, Research Group ExpORL, 3000 Leuven, Belgium}
\affil[3]{Aarhus University, Department of Electrical and Computer Engineering, Center for Ear-EEG, 8200 Aarhus N, Denmark}

\affil[*]{simon.geirnaert@kuleuven.be}

\begin{abstract}
	Current assistive hearing devices, such as hearing aids and cochlear implants, lack the ability to adapt to the listener's focus of auditory attention, limiting their effectiveness in complex acoustic environments like cocktail party scenarios where multiple conversations occur simultaneously. Neuro-steered hearing devices aim to overcome this limitation by decoding the listener’s auditory attention from neural signals, such as electroencephalography (EEG). While many auditory attention decoding (AAD) studies have used high-density scalp EEG, such systems are impractical for daily use as they are bulky and uncomfortable. Therefore, AAD with wearable and unobtrusive EEG systems that are comfortable to wear and can be used for long-term recording are required. Around-ear EEG systems like cEEGrids have shown promise in AAD, but in-ear EEG, recorded via custom earpieces offering superior comfort, remains underexplored. We present a new AAD dataset with simultaneously recorded scalp, around-ear, and in-ear EEG, enabling a direct comparison. Using a classic linear stimulus reconstruction algorithm, a significant performance gap between all three systems exists, with AAD accuracies of 83.4\% (scalp EEG), 67.2\% (around-ear EEG), and 61.1\% (in-ear EEG) on 60s decision windows. These results highlight the trade-off between decoding performance and practical usability. Yet, while the ear-based EEG systems using basic algorithms might currently not yield accurate enough performances for a decision speed-sensitive application in hearing aids, their significant performance suggests potential for attention monitoring on longer timescales. Furthermore, adding an external reference or a few scalp electrodes via greedy forward selection substantially and quickly boosts accuracy by over 10 percent point, especially for in-ear EEG. These findings position in-ear EEG as a promising component in EEG sensor networks for AAD.
\end{abstract}

\begin{document}
	
	\flushbottom
	\maketitle
	%
	%
	\thispagestyle{empty}
	
	\section*{Introduction}
	Neuro-steered hearing devices allow hearing aid and cochlear implant users to cognitively steer their assistive device towards the conversation they want to listen to~\cite{osullivan2014attentional,geirnaert2021eegBased}. Current assistive hearing devices lack a fundamental piece of information about the listener's attentional focus in cocktail party scenarios, where multiple conversations occur simultaneously. This prevents the hearing device from correctly identifying and enhancing the conversation of interest, while suppressing other unattended conversations and noise sources. Assistive hearing devices that can identify the attended speaker by decoding the user's auditory attention could therefore greatly improve the quality of life for people with hearing impairments.
	
	Crucial to enabling neuro-steered hearing devices is the ability to decode auditory attention from the brain signals of the listener. In this context, electroencephalography (EEG) is often the neuro-recording modality of choice, as it is non-invasive, wearable, and relatively cheap~\cite{nicolasAlonso2012brain}, all critical factors when considering integration into widespread, daily-use hearing devices. Furthermore, the high temporal resolution of EEG facilitates easy tracking of auditory attention to speech. 
	
	EEG-based auditory attention decoding (AAD) algorithms generally fall into two categories~\cite{geirnaert2021eegBased}. The first category relies on neural tracking, i.e., the neural signals of the listener entrain more strongly with features of the attended speech signal than with unattended ones~\cite{mesgarani2012selective,ding2012emergence,osullivan2014attentional}. These features often include acoustic ones such as the envelope or spectrogram~\cite{mesgarani2012selective,ding2012emergence,osullivan2014attentional}, but also linguistic features as phoneme surprisal or semantic dissimilarity are used~\cite{diLiberto2015low,gillis2021neural}. The second category aims to directly classify attention from EEG signals, for example based on the spatial direction of attention \cite{geirnaert2020fast,vandecappelle2021eeg,geirnaert2021riemannian}, or using more black-box approaches based on deep learning~\cite{geirnaert2021eegBased,puffay2023relating,nguyen2024aadnet}. However, while the first category of algorithms based on neural tracking have consistently shown robustness across a wide range of datasets and setups, the second category of direct classification approaches are much more prone to overfitting, shortcut learning, and overestimated performance~\cite{geirnaert2021eegBased,puffay2023relating,rotaru2023what,ivucic2024impact,yan25overestimated}. For this reason, we only consider the first category of algorithms based on neural tracking in this study.
	
	Many of these AAD algorithms are developed using high-density, full-scalp EEG datasets, several of which are publicly available. While these high-density, wet EEG systems are well-suited for developing new AAD algorithms, they are too bulky and impractical for daily-life usage in neuro-steered hearing devices. In practice, such devices require wearable, concealable, and dry EEG sensor systems that are comfortable to wear and capable of long-term recording. However, these wearable and concealable EEG systems typically include fewer electrodes with limited spatial coverage. This makes it crucial to evaluate and develop AAD algorithms specifically for such systems, taking all these constraints into account.
	
	Research towards wearable EEG systems for AAD generally takes two distinct avenues: (1) a data-driven determination of optimal electrode locations based on high-density EEG, assuming ideal miniaturized EEG sensors are available, and (2) a recording system-driven approach, starting from existing wearable EEG recording systems. The data-driven approach assumes the existence of miniaturized EEG sensor nodes that can be flexibly mounted at optimally selected locations. Using data-driven channel selection and reduction techniques, Mirkovic et al.~\cite{mirkovic2015decoding} (using backward elimination) and Mundanad Narayanan and Bertrand~\cite{narayanan2020analysis} (using a greedy utility-based selection) were able to reduce the number of EEG channels used for neural tracking-based algorithms from 96 to 25 and 64 to 10, respectively, without a loss in AAD performance. Furthermore, Mundanad Narayanan et al.~\cite{narayanan2021eeg} were able to reduce inter-electrode distances to $\SI{3}{\centi\meter}$, eliminating the need for long-distance montages and thereby increasing the practicality of wireless EEG sensor networks for AAD. These data-driven channel selection methods, all starting from neural tracking, typically select, as expected, electrodes above the temporal lobe, where the auditory cortex is located.
	
	The second, more practical approach starts from existing wearable EEG systems that are already available. While several efforts have aimed to make traditional scalp EEG more portable (for example, by making systems wireless and using portable amplifiers), also in the context of AAD~\cite{ciccarelli2019comparison,straetmans2021neural,ha2023validation,straetmans2024neural,hjortkjaer2025real}, such systems might still be too obtrusive and uncomfortable for long-term, daily-life use, as envisioned in neuro-steered hearing devices. As a result, the focus here goes to ear-based EEG systems, which are easier to integrate with hearing devices and are recorded at an interesting location, i.e., near the temporal lobe. These systems enable wearable, unobtrusive, and long-term EEG monitoring while maintaining user comfort. Two configurations exist: around-ear and in-ear EEG. Around-ear systems, such as cEEGrids~\cite{debener2015unobtrusive,bleichner2017concealed}, typically position electrodes in a flexible array behind the ear, and have been used in various wearable EEG applications, including sleep monitoring~\cite{mikkelsen2019machine}, epilepsy seizure detection~\cite{gu2017comparison}, and mental load monitoring~\cite{cretot2023assessing}. In-ear systems, which place electrodes inside the ear using an earpiece~\cite{looney2012in,mikkelsen2015eeg,kappel2019dry}, have also shown successful in applications such as sleep monitoring~\cite{mikkelsen2019accurate} and hearing assessment~\cite{christensen2018toward}. These in-ear EEG systems offer specific advantages as they are highly unobtrusive and concealable, and comfortable for prolonged wear~\cite{looney2012in}. Although many of these systems still require wired connections - for example, to an amplifier -, wireless, miniaturized EEG sensor nodes, also for ear-based recordings, are currently being developed~\cite{ding2025synchronized}.
	
	In the context of AAD, around-ear EEG has repeatedly demonstrated success with neural tracking algorithms. In Mirkovic et al.~\cite{mirkovic2016target}, a two-speaker scenario yielded an accuracy of $69.3\%$ using around-ear EEG, compared to $84.8\%$ with high-density scalp EEG, on $\SI{60}{\second}$ windows. Comparable results were reported in Holtze et al.~\cite{holtze2022ear}. While lower accuracies were found by Nogueira et al.\cite{nogueira2019decoding}, they nonetheless demonstrated that AAD with around-ear EEG is feasible in cochlear implant users. In a more complex four-speaker scenario, Zhu et al.~\cite{zhu2025using} reported an AAD accuracy of $41.3\%$ on $\SI{60}{\second}$ windows with around-ear EEG versus $75.5\%$ with scalp EEG. In-ear EEG for AAD was initially explored in Fiedler et al.~\cite{fiedler2017single}, using three in-ear electrodes per ear. In a small dataset of four participants and using a forward model (predicting EEG channels from the stimulus feature), significant performance was only achieved using an external scalp reference electrode (FT7). Similarly, Thornton et al.~\cite{thornton2024comparison} used two in-ear electrodes per ear, referenced to an external scalp reference (FT7), and applied both linear and CNN-based neural tracking algorithms. On a larger dataset of 18 participants, they found modest yet significant AAD accuracies around $64\%$ on $\SI{30}{\second}$ windows. Notably, the CNN-based algorithm did not outperform linear models (with even indications of it performing worse), which the authors hypothesized is due to the lower signal-to-noise ratio (SNR) of in-ear EEG. This lower SNR could cause the CNN to overfit to artifacts, something linear models seem less prone to. Therefore, in this study, we adopt a linear model, furthermore facilitating an interpretable direct comparison between EEG setups. Importantly, both studies relied on an external scalp reference electrode. While such a setup may still be practical, it does not constitute a truly `ear-based' EEG system that can be seamlessly integrated with hearing devices. Lastly, Nguyen et al.~\cite{nguyen2025cognitive} adopted a different approach to tackle the AAD problem with in-ear EEG, using event-related potentials and showing attentional modulation of the cognitive brain response to specific target words.
	
	However, to date, no study in the AAD literature has directly compared scalp, around-ear, and in-ear EEG. Furthermore, as noted above, in-ear AAD studies have consistently relied on an external scalp reference. To address this, we introduce a new AAD dataset in which participants attended to one of two competing talkers while their scalp, around-ear, and in-ear EEG is simultaneously recorded. This allows for a direct comparison of AAD performance across all three systems on the same data, and enables the development of tailored preprocessing pipelines and AAD algorithms for ear-based EEG systems - for example, leveraging transfer learning between scalp and ear-EEG. In this paper, we present the newly recorded AAD ear EEG-based dataset, compare the different EEG systems using a standard neural tracking algorithm, and investigate system complementarity, the influence of the reference electrode, and the potential of constructing an EEG sensor network centered around ear-based systems. To facilitate future development of ear-based AAD algorithms, the dataset is made publicly available~\cite{geirnaert2025ear}.  
	
	\section*{Dataset}
	
	In this section, we provide a full description of the protocol and experimental setup for the new AAD dataset with simultaneous scalp, around-ear, and in-ear EEG recordings. The study was approved by the Institutional Review Board at Aarhus University (approval number 2024-0673174) and conducted following the applicable guidelines and regulations. The full dataset, along with a detailed manual and description, is publicly available~\cite{geirnaert2025ear}.
	
	\subsection*{Participants}
	15 adults with self-reported normal hearing (seven women, eight men) participated in the study. Participants were between 19 and 31 years old (mean age: 24.7 years, standard deviation: 3.5). All were native Danish speakers and provided written informed consent in accordance with the approved protocol.
	
	\subsection*{Protocol and stimuli}
	
\begin{figure}
		\centering
		\includegraphics[width=\linewidth]{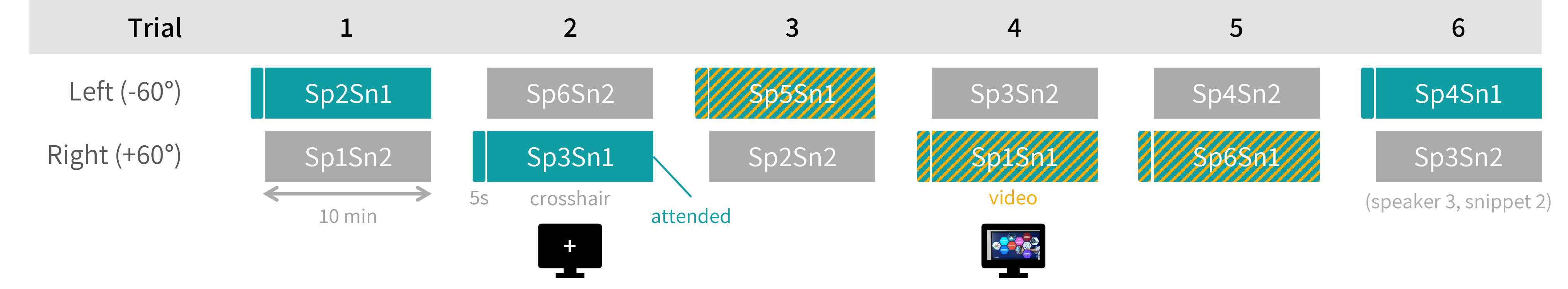}
		\caption{The protocol consists of six $\SI{10}{\minute}$-trials per participant, each involving attention to one of two competing speech signals (left and right) from six different speakers. A $\SI{5}{\second}$ head start with only the target stream precedes each trial. In half of the trials, a video of the attended speaker is shown (shaded), while in the other half, a fixation cross is displayed. Attention direction is balanced across trials.}
		\label{fig:protocol}
	\end{figure}

	Each participant took part in a selective auditory attention experiment, in which they were instructed to attended to one of two simultaneously presented speech streams (left or right) while seated in a room without electric shielding. The experiment consisted of six trials, each lasting $\SI{10}{\minute}$, resulting in a total of $\SI{60}{\minute}$ of data per participant. The two competing speech signals were presented via insert earphones, integrated into the custom earpieces used for in-ear EEG recordings (see below), and were spatially simulated at $\pm 60^{\circ}$ azimuth using head-related transfer functions. The signals were normalized to $-33$ loudness units relative to full scale (LUFS). At the beginning of each trial, participants were presented a $\SI{5}{\second}$ prelisten of the to-be-attended speech stream, followed by $\SI{10}{\minute}$ of competing speech signals. This prelisten fragment was intended to help participants identify the correct target stream. Figure~\ref{fig:protocol} illustrates the experimental protocol. After each trial, participants answered an open-ended comprehension question requiring a short, precise answer about the content of the attended speech. Performance on these questions was not logged, nor used as an exclusion criterion, but was only included to help maintain the participant’s attention and motivation. Breaks were allowed between trials at the participants’ discretion.
	
	The speech material consisted of six educational talks aimed at high school students, delivered by six different male Danish professors at Aarhus University. Both audio and video were available for each talk. From each talk, two non-overlapping $\SI{10}{\minute}+\SI{5}{\second}$ excerpts were extracted and assigned to the set of attended and unattended stimuli, respectively, with each excerpt used exactly and only once during the experiment. In total, there were thus six different speakers to attend to, reducing potential bias due to familiarization with a specific speaker or voice. Attended excerpts were chosen such that they could be fully understood without prior context or knowledge. Unlike many other AAD studies that truncate silences to a maximal duration, the speech signals here were not further manipulated, i.e., silences were left at their natural length. Across all stimuli, silences longer than $\SI{250}{\milli\second}$ had a mean ($\pm$ standard deviation) duration of $\SI{667}{\milli\second} \pm \SI{455}{\milli\second}$. As silence duration is known to affect onset processing and neural tracking correlations of envelopes~\cite{deoisres2023continuous}, this difference should be considered when comparing with other AAD studies. Full details on the talks and extracted excerpts are available online~\cite{geirnaert2025ear}.
	
	For each participant, in half of the trials, the video of the attended speaker was displayed on a screen in front of the participant. In the other half, the participants were instructed to focus on a fixation cross displayed in the center of the screen. The direction of attention (left or right) was also counterbalanced across all six trials. The pairing of the attended and unattended $\SI{10}{\minute}$-snippets was done such that the same speaker never appeared simultaneously in the attended and unattended snippet. This pairing of attended/unattended snippets, the order of attended snippets (and thus trials), attended direction (left/right), and video/no video conditions were randomized per participant. Across participants, care was taken to balance left/right attention direction for video/no-video conditions, to avoid a confound between video viewing and attended direction. Similarly, care was taken to ensure that each video was viewed approximately the same number of times across participants. Triggers were used to synchronize the audio and video with the EEG data. Audio triggers were generated from a separate audio trigger channel and a photosensor in the top right corner of the screen was used to generate triggers from a flashing box in the video stream.
	
	At the end of the experiment, an additional $\SI{5}{\minute}$ auditory steady-state response (ASSR) recording using a broadband chirp stimulus was conducted to assess SNR across the EEG setups. This paper does not report further analysis of the ASSR data.
	
	\subsection*{EEG data acquisition}
\begin{figure}
		\centering
		\begin{subfigure}{0.5\linewidth}
			\centering
			\includegraphics[width=1\linewidth]{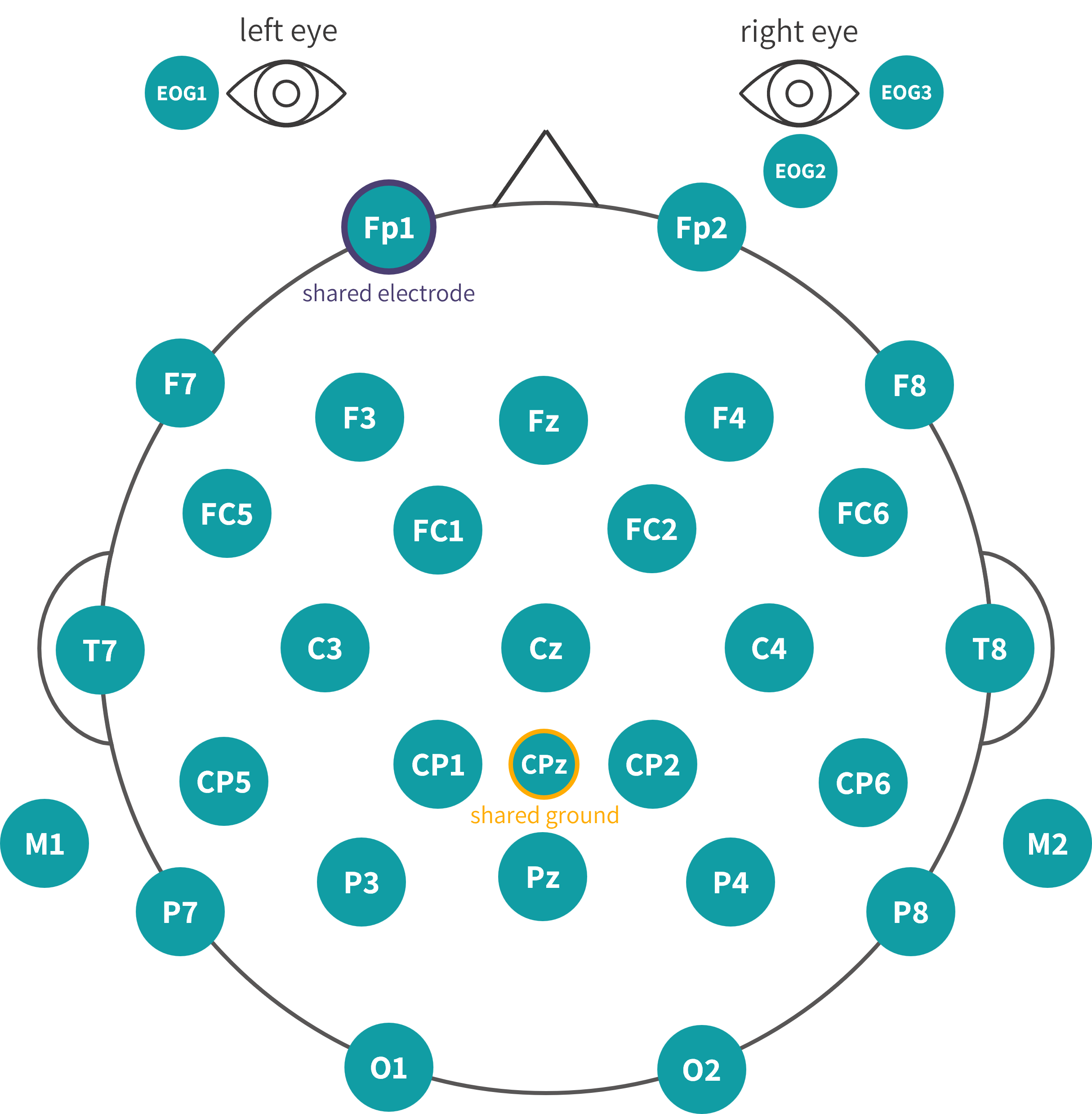}
			\caption{}
			\label{fig:scalp-setup}
		\end{subfigure}
		
		\begin{subfigure}{0.5\linewidth}
			\centering
			\includegraphics[width=0.75\linewidth]{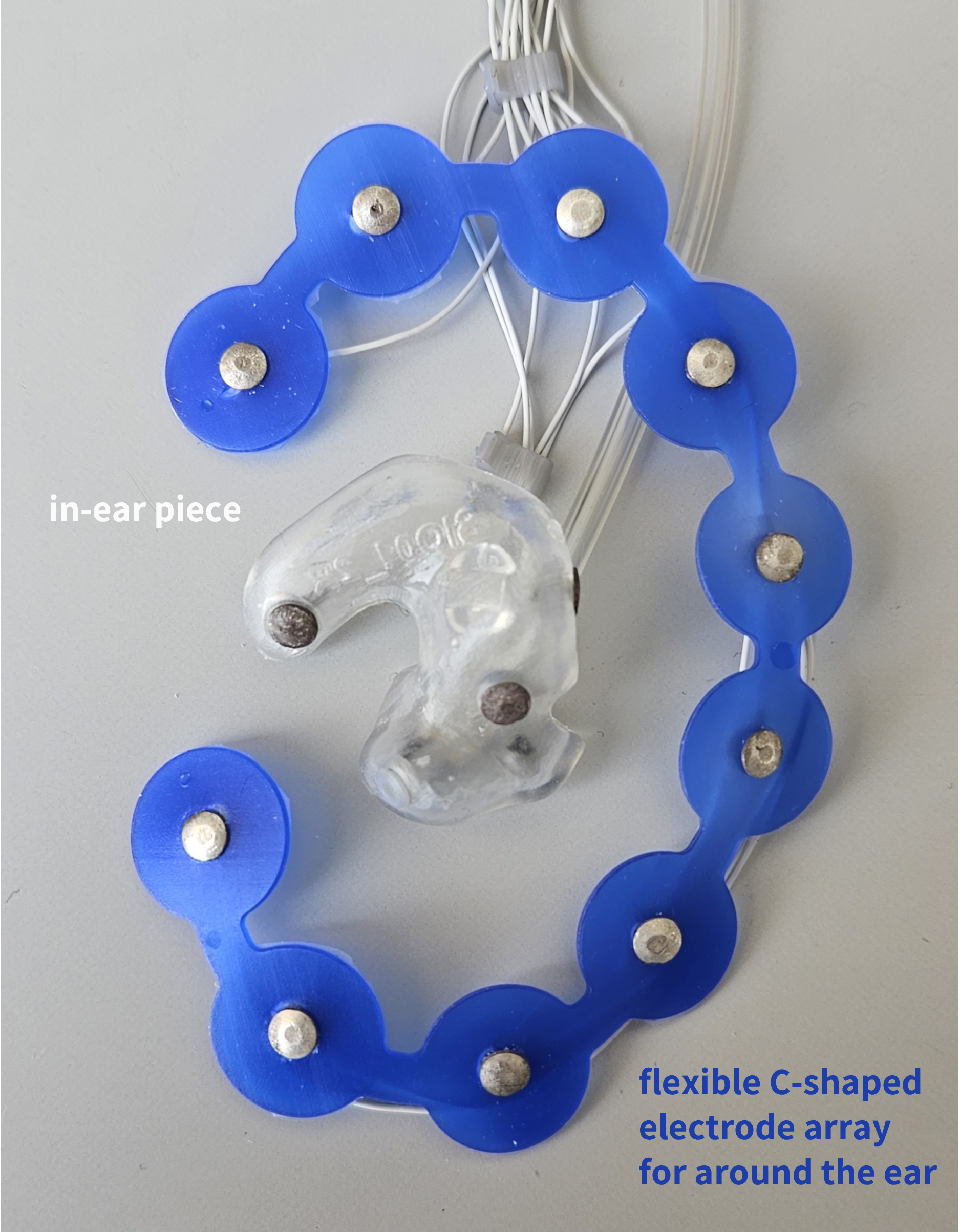}
			\caption{}
			\label{fig:around+in-ear-setup}
		\end{subfigure}%
		\begin{subfigure}{0.5\linewidth}
			\centering
			\includegraphics[width=0.9\linewidth]{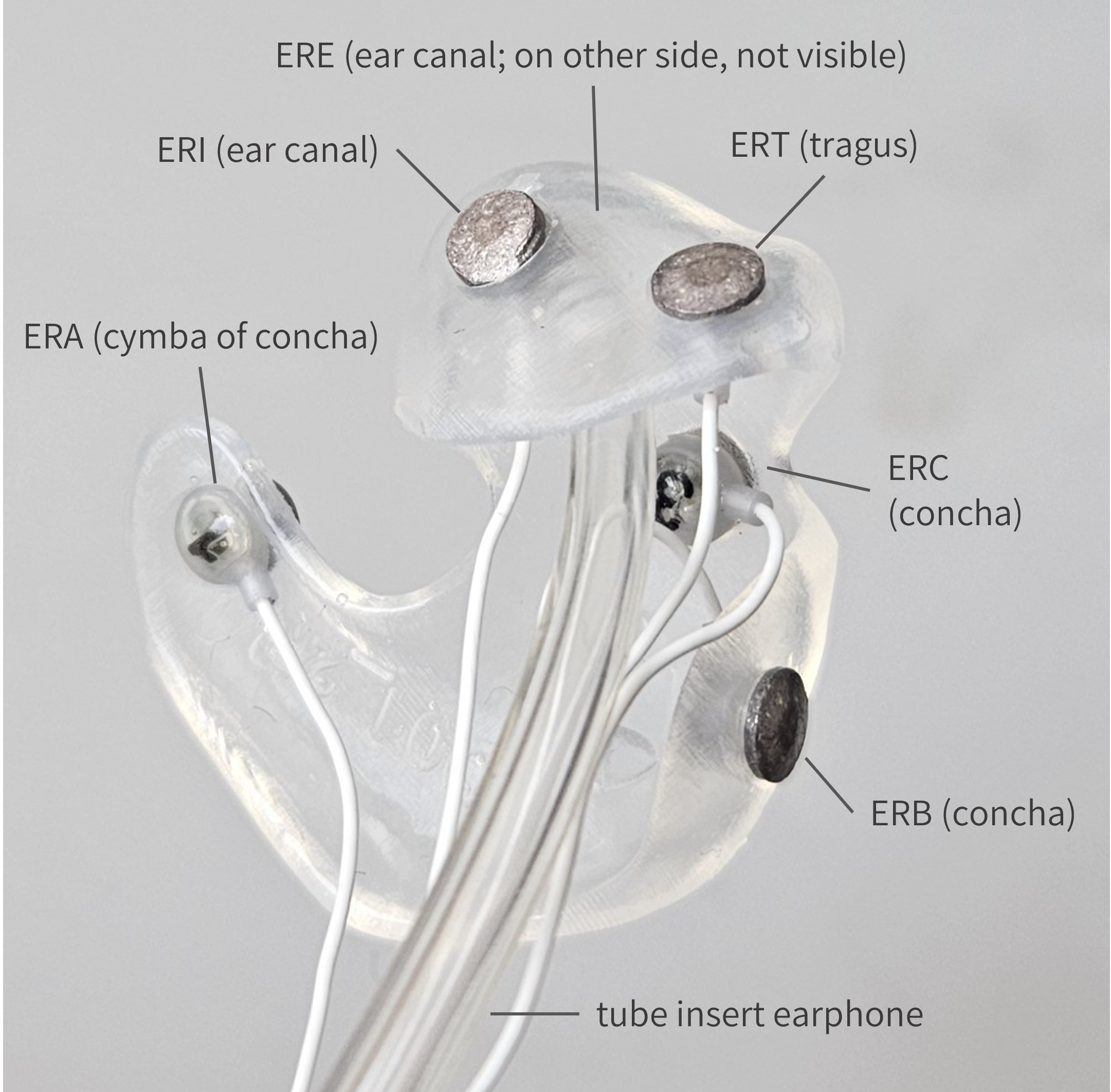}
			\caption{}
			\label{fig:in-ear-setup}
		\end{subfigure}%
		\caption{\textbf{(a)} The scalp EEG setup with 32 electrodes, including three for EOG, with Fp1 a shared electrode position between the scalp and ear-EEG system, and CPz the shared ground electrode position. \textbf{(b)} The 3D-printed flexible C-shaped array for around-ear EEG, and a customized earpiece for in-ear EEG. \textbf{(c)} The in-ear electrode labeling and positions in an earpiece for a specific participant, here for the right ear.}	
		\label{fig:eeg-setup}
	\end{figure}

	Scalp, around-ear, and in-ear EEG were recorded simultaneously during the selective auditory attention task using two TMSi Mobita amplifiers (sampling rate: $\SI{1000}{\hertz}$), each allowing to record EEG from 32 electrodes. One amplifier was used for scalp EEG and one for ear-EEG (including both around-ear and in-ear EEG). To enable merging of the two EEG data streams in the post-processing analysis, both systems had an electrode at a shared location at the Fp1 position, allowing to bring both EEG setups in the same `reference' system. Similarly, both systems used a ground electrode at a shared location, i.e., CPz. The EEG setup is summarized in Table~\ref{tab:eeg-setup} and visualized in Figure~\ref{fig:eeg-setup}.
	
	\begin{table}
		\centering
		\caption{Overview of the EEG recording setup.}
		\label{tab:eeg-setup}
		\renewcommand{\arraystretch}{1.5}
		\begin{tabular}{@{}lp{1.4cm}p{2.2cm}p{1.8cm}p{4cm}p{3.6cm}@{}}
			\toprule
			\textbf{EEG setup} & \textbf{Amplifier} &  \textbf{Number of} \newline \textbf{electrodes} & \textbf{Electrodes} & \textbf{Electrode placement} & \textbf{Notes} \\ \midrule
			Scalp EEG         & TMSi \newline Mobita 1 & 32 \newline(incl. 3 EOG)              & Wet circular \newline Ag/AgCl, \diameter $\SI{4}{\milli\meter}$           & International 10-20 system; see Figure~\ref{fig:scalp-setup} & Cap mounted over ear-EEG setup; ground electrode at CPz position \\
			Around-ear EEG    & TMSi \newline Mobita 2 & 19 \newline(9 left, 10 right)         & Wet circular \newline Ag/AgCl, \diameter $\SI{4}{\milli\meter}$            & Around the ear using flexible 3D-printed array (cEEGrid-style); see Figure~\ref{fig:around+in-ear-setup} & Ground electrode at CPz position \\
			In-ear EEG        & TMSi \newline Mobita 2 & 12 \newline(6 per ear)                & Dry circular \newline Ag/AgCl, \diameter $\SI{4}{\milli\meter}$             & 2 in ear canal (E, I), 1 at tragus (T), 3 in concha region (A, B, C)~\cite{kidmose2013study}; Figure~\ref{fig:in-ear-setup}; exact positioning varied between participants due to anatomical differences & Soft, silicone-based earpieces individualized to match the unique anatomy of each ear~\cite{kappel2019dry}; same amplifier as around-ear EEG \\
			Fp1-electrode     & TMSi \newline Mobita 2 & 1                & Wet circular \newline Ag/AgCl, \diameter $\SI{4}{\milli\meter}$             & See Figure~\ref{fig:scalp-setup} & Additional scalp electrode to provide shared reference with scalp EEG \\
			\bottomrule
		\end{tabular}
		\renewcommand{\arraystretch}{1}
	\end{table}
	
	\subsubsection*{Scalp EEG}
	Scalp EEG was recorded using 32 wet circular Ag/AgCl-electrodes with a diameter of $\SI{4}{\milli\meter}$~\cite{kappel2022characterization}, of which 29 were placed according to the international 10-20 system (Figure~\ref{fig:scalp-setup}) and three were used to record eye activity (EOG). The scalp EEG cap was mounted on top of the ear-EEG setup.
	
	\subsubsection*{Around-ear EEG}
	Around-ear EEG was recorded using 19 Ag/AgCl-electrodes~\cite{kappel2022characterization} (the same type of the scalp EEG), with 9 placed around the left ear and 10 around the right ear. The electrodes were positioned behind and around the ear using a flexible, 3D-printed C-shaped array, similar to cEEGrids~\cite{debener2015unobtrusive}, and were affixed using adhesive stickers (Figure~\ref{fig:around+in-ear-setup}). One electrode at the bottom of the left array (near the cheek) was sacrificed to provide the shared electrode at the Fp1-position. To maximize signal quality, special care was taken during application (e.g., moving hair out of the way, skin cleaning), following best practices for around-ear EEG outlined in H{\"o}lle and Bleichner~\cite{holle2023recording}. Additionally, a small droplet of conductive gel was applied to each electrode.
	
	\subsubsection*{In-ear EEG}
	In-ear EEG was recorded using 12 dry Ag/AgCl-electrodes~\cite{kappel2022characterization} (the same type as those used for scalp and around-ear EEG), six per ear. For each participant, individually modeled earpieces were fabricated from flexible silicone material to ensure optimal comfort and electrode-skin contact. Although all earpieces were customized to match the anatomical shape of each participant's ear, they followed a common global design, as shown in Figure~\ref{fig:in-ear-setup} and described in Kappel et al.~\cite{kappel2019dry}. 
	
	Electrodes were placed on regions of the earpiece with a large radius of curvature, allowing them to follow the overall shape of the earpiece while still protruding slightly from its surface. Placement followed the system and naming convention described by Kidmose et al.~\cite{kidmose2013study}. Specifically, electrodes were placed in the ear canal (two electrodes: ExI and ExE, with “x” denoting the ear side), at the tragus (one electrode: ExT), and in the concha area (three electrodes: ExA, ExB, and ExC), as illustrated in Figure~\ref{fig:in-ear-setup}. Due to anatomical variability, exact electrode locations varied slightly between participants.
	
	Importantly, unlike the scalp and around-ear electrodes, the in-ear electrodes were used dry, i.e., without electrode gel. The ears were cleaned prior to insertion using a damp cotton swab.
	
	\section*{Methods}
	
	\subsection*{Auditory attention decoding algorithm based on neural tracking}
	
	We adopt a classic linear stimulus reconstruction approach for AAD, where features of the attended speech signal are reconstructed from the listener’s EEG~\cite{osullivan2014attentional,geirnaert2021eegBased}. Due to neural tracking, the reconstruction should correlate more strongly with the attended than the unattended speech features~\cite{mesgarani2012selective,ding2012emergence}. The attended speaker is identified as the one whose speech features yield the highest Pearson correlation with the reconstruction. While non-linear (e.g., CNN-based) and canonical correlation analysis-based methods exist, they do not consistently outperform this approach~\cite{geirnaert2021eegBased}, and especially not on ear-based EEG~\cite{thornton2024comparison}, justifying our choice for the linear stimulus reconstruction algorithm. We here give a short explanation of the algorithm; extensive descriptions of this algorithm can be found in literature~\cite{osullivan2014attentional,geirnaert2021eegBased,geirnaert2022timeAdaptive}.
	
	Given $T$ samples of time-lagged EEG $\mat{X} \in \R^{T \times CL}$ with $C$ channels and $L$ time lags, and the one-dimensional attended speech feature $\vec{s}_{\text{a}} \in \R^T$, the goal is to find a backward decoder $\vec{d} \in \R^{CL}$ that optimally reconstructs the attended speech features from the EEG. This decoder represents a spatio-temporal filter as it linearly integrates EEG across all $C$ channels and $L$ time lags to find the reconstruction $\hat{\vec{s}}_{\text{a}} = \mat{X}\vec{d}$. The time lags are typically taken after the current time sample, as the neural response follows after the stimulus. In this setup, $\mat{X} = \begin{bmatrix}
		\mat{X}_1 & \dots & \mat{X}_C \end{bmatrix}$ is a block Hankel matrix with per-channel Hankel matrices $\mat{X}_c \in \R^{T \times L}$ containing the time-lagged EEG data of the $c$\textsuperscript{th} channel:
	\[ \mat{X}_{c} = \begin{bmatrix}
		x_{c}(0) & x_{c}(1) & \cdots & x_{c}(L-1) \\
		x_{c}(1) & x_{c}(2) & \cdots & x_{c}(L) \\
		\vdots & \vdots &  & \vdots \\
		x_{c}(T-1) & 0 & \cdots & 0\\
	\end{bmatrix},
	\]
	with $x_c(t)$ the EEG sample at the $t$\textsuperscript{th} time sample index and $c$\textsuperscript{th} channel. In this paper, we take $L$ such that the filter taps span $[0,400]\SI{}{\milli\second}$ post-stimulus~\cite{geirnaert2022timeAdaptive}.
	
	The optimal reconstruction is defined by minimizing the squared error with the attended speech feature (which is equivalent to maximizing the Pearson correlation~\cite{biesmans2017auditory}), including an $L_2$-norm regularization term with hyperparameter $\lambda$ to prevent overfitting:
	\begin{equation}
		\label{eq:ls}
		\hat{\vec{d}} = \underset{\vec{d}}{\text{argmin}} \norm{\vec{s}_{\text{a}}-\hat{\vec{s}}_{\text{a}}}_2^2 + \lambda\norm{\vec{d}}_2^2 = \underset{\vec{d}}{\text{argmin}} \norm{\vec{s}_{\text{a}}-\mat{X}\vec{d}}_2^2+ \lambda\norm{\vec{d}}_2^2.
	\end{equation}
	The normal equations then give the solution to equation~\eqref{eq:ls}:
	\[
	\hat{\vec{d}} = \inverse{\left(\transpose{\mat{X}}\mat{X}+\lambda \mat{I}\right)}\transpose{\mat{X}}\vec{s}_{\text{a}}.
	\]
	While the hyperparameter $\lambda$ is often set via cross-validation, we adopt the analytical estimator by Ledoit and Wolf~\cite{ledoit2004well}, which has shown good performance for this AAD decoder~\cite{geirnaert2021unsupervised,geirnaert2022timeAdaptive}. During training, the attended speech feature $\vec{s}_{\text{a}}$ is assumed known to be able to solve equation~\eqref{eq:ls}. When attention labels are unavailable, an unsupervised training algorithm can be applied~\cite{geirnaert2021unsupervised,geirnaert2022timeAdaptive}.

	To determine attention at test time for a new EEG segment $\mat{X}^{(\text{test})} \in \R^{T_\text{(test)} \times CL}$ with $T_{\text{test}}$ samples (the decision window length) and speech features $\vec{s}_1^{\text{(test)}}$ and $\vec{s}_2^{\text{(test)}}$ from the two competing speakers, first, a reconstruction is made from the EEG using the learned decoder: $\hat{\vec{s}}^{\text{(test)}}_{\text{a}} = \mat{X}^{\text{(test)}}\hat{\vec{d}}$. The attended speaker is identified as the one whose feature yields the highest Pearson correlation with the reconstruction $\left(\text{corr}\!\left(\hat{\vec{s}}^{\text{(test)}}_{\text{a}},\vec{s}_1^{\text{(test)}}\right) \text{ vs. } \text{corr}\!\left(\hat{\vec{s}}^{\text{(test)}}_{\text{a}},\vec{s}_2^{\text{(test)}}\right)\right)$.
	
	\subsection*{Preprocessing}
	
	\subsubsection*{EEG preprocessing}
	Scalp, around-ear, and in-ear EEG signals follow the same preprocessing steps, yet they are preprocessed separately (there is never information used from, for example, scalp EEG in the preprocessing of in-ear EEG). First, EEG data is filtered between $\SIrange{1}{9}{\hertz}$ using a zero-phase bandpass filter with a $4$\textsuperscript{th}-order Butterworth polynomial. This frequency band has shown successful multiple times with acoustic speech features like the envelope~\cite{osullivan2014attentional,mirkovic2015decoding,geirnaert2021eegBased}. Then, EEG data is split into the $\SI{10}{\minute}$-trials based on the trigger signals. All following preprocessing steps are therefore applied per trial. 
	
	For scalp EEG, eye artifacts are removed by regressing out EOG channels using least squares. As we consider the EOG channels to be part of the high-density scalp EEG system, this eye artifact removal is omitted for around-ear and in-ear recordings. Bad channels (correlation on $\SI{2}{\second}$ segments below $0.45$ for at least half the trial with $95\%$ of the other channels) are subsequently removed using EEGLAB's \textit{clean\_channels}-function~\cite{delorme2004eeglab}. Artifact subspace reconstruction~\cite{chang2020evaluation} (ASR) is then applied to remove and reconstruct (high-power) artifactual segments using EEGLAB's \textit{clean\_asr}-function (cutoff 5, default settings)~\cite{delorme2004eeglab}. Any remaining high-power artifacts are removed using a modified version of EEGLAB's \textit{clean\_windows}-function, identifying outlying RMS-powers per $\SI{1}{\second}$-window across the whole channel, and subsequently removing bad windows per channel individually. Bad channels and time segments are excluded during re-referencing and normalization, and are afterwards replaced with zeros to be handled by the data-driven filtering. 
	
	Lastly, the data are downsampled to $\SI{20}{\hertz}$ (without anti-aliasing filtering to not introduce additional filtering artifacts) and normalized per trial such that the Frobenius norm across time and channels is $1$.
	
	By default, common average referencing (CAR) is applied per EEG setup, i.e., re-referencing to the average of all channels per setup. The shared Fp1-electrode is excluded in this CAR re-referencing for the around-ear and in-ear data, and is only used to re-reference those setups when they are combined with scalp EEG electrodes.
	
	\subsubsection*{Speech feature extraction}
	The acoustic envelope, representing the amplitude modulation of the speech signal, is used as the speech feature that is reconstructed from the EEG, a common choice in neural tracking-based AAD~\cite{osullivan2014attentional,mirkovic2015decoding,geirnaert2021eegBased}. Although prior AAD studies using in-ear EEG~\cite{fiedler2017single,thornton2024comparison} adopted the onset envelope as an (additional) feature, we found no added value.
	
	Envelopes are extracted using the procedure proposed in Biesmans et al.~\cite{biesmans2017auditory}, involving decomposition with a gammatone filterbank that mimics human auditory processing (19 bands with center frequencies between $\SI{50}{\hertz}$ and $\SI{5}{\kilo\hertz}$), powerlaw compression per subband (exponent $0.6$), and summation into one envelope. The resulting envelope is zero-phase filtered between $\SIrange{1}{9}{\hertz}$ using a $4$\textsuperscript{th}-order Butterworth to match it spectrally with the EEG, and is downsampled to $\SI{20}{\hertz}$. Per trial, the envelopes are then normalized to zero mean and unit variance.
	
	\subsection*{Evaluation procedure}
	We use leave-one-trial-out cross-validation (CV) for participant-specific decoding (i.e., when neural decoders are trained for a specific participant), with six folds per participant, each excluding a full $\SI{10}{\minute}$ trial. While such a strict CV procedure generally results in slightly lower performances than when mixing data from within a trial in both training and test set, several papers have shown the importance of a strict validation procedure in AAD to avoid overfitting and exploitation of shortcuts~\cite{puffay2023relating,rotaru2023what,ivucic2024impact}. While this is less of an issue with neural tracking-based AAD algorithms as used here, we find it important to adhere to a stricter validation procedure as a more general guideline in AAD and machine learning research. Furthermore, the leave-one-trial-out CV guarantees that no information leakage from the training to the test set is possible due to the per-trial preprocessing. For participant-independent decoding, a leave-one-participant-out CV procedure is used.
	
	AAD accuracy is evaluated for varying decision window lengths $T_{\text{test}} \in \{1,5,10,30,60,120,300,600\} \SI{}{\second}$. As AAD accuracy for correlation-based neural tracking algorithms declines with shorter windows~\cite{geirnaert2020interpretable}, we report the full accuracy vs. decision window length performance curve. When a single decision window length is required for comparison, we use $\SI{60}{\second}$ as it is long enough to reduce correlation estimation noise, while still providing sufficient decisions per participant.
	
	To summarize performance across this curve, we compute the minimal expected switch duration (MESD)~\cite{geirnaert2020interpretable} (in seconds or minutes), which determines the optimal trade-off between accuracy and decision window length by quantifying the expected time needed to switch speaker gain after an attention switch. Lower MESD (better) indicates the algorithm can facilitate attention switches faster. 
	
	The significance level of AAD accuracy is determined using the inverse binomial cumulative distribution function with the number of decision windows and a $50\%$ success probability as parameters, evaluated at $0.95$ ($\alpha$-level $0.05$). When assessing the AAD accuracy per individual participant, the number of decision windows corresponds to $\frac{\SI{60}{\minute} \times \SI{60}{\second}/\SI{}{\minute}}{T_{\text{test}}}$. When assessing the average AAD accuracy across participants, the number of decision windows is correspondingly multiplied by the number of participants ($15$). 
	
	\section*{Results and discussion}
	\begin{figure}
		\centering
		\begin{subfigure}{1\linewidth}
			\centering
			\includegraphics[height=0.35\textheight]{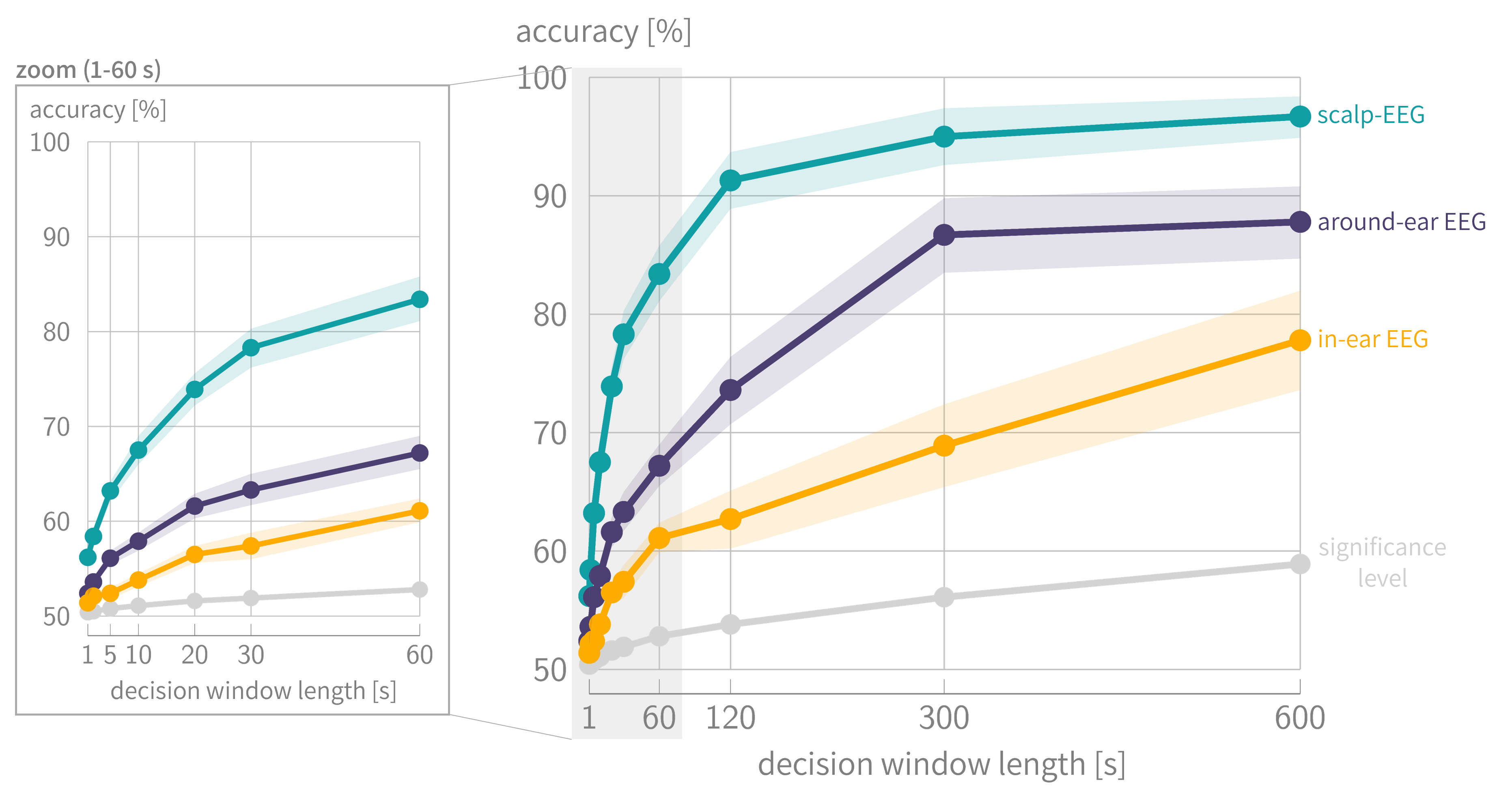}
			\caption{}
			\label{fig:baseline-comparison-car}
		\end{subfigure}
		
		\begin{subfigure}{1\linewidth}
			\centering
			\includegraphics[height=0.275\textheight]{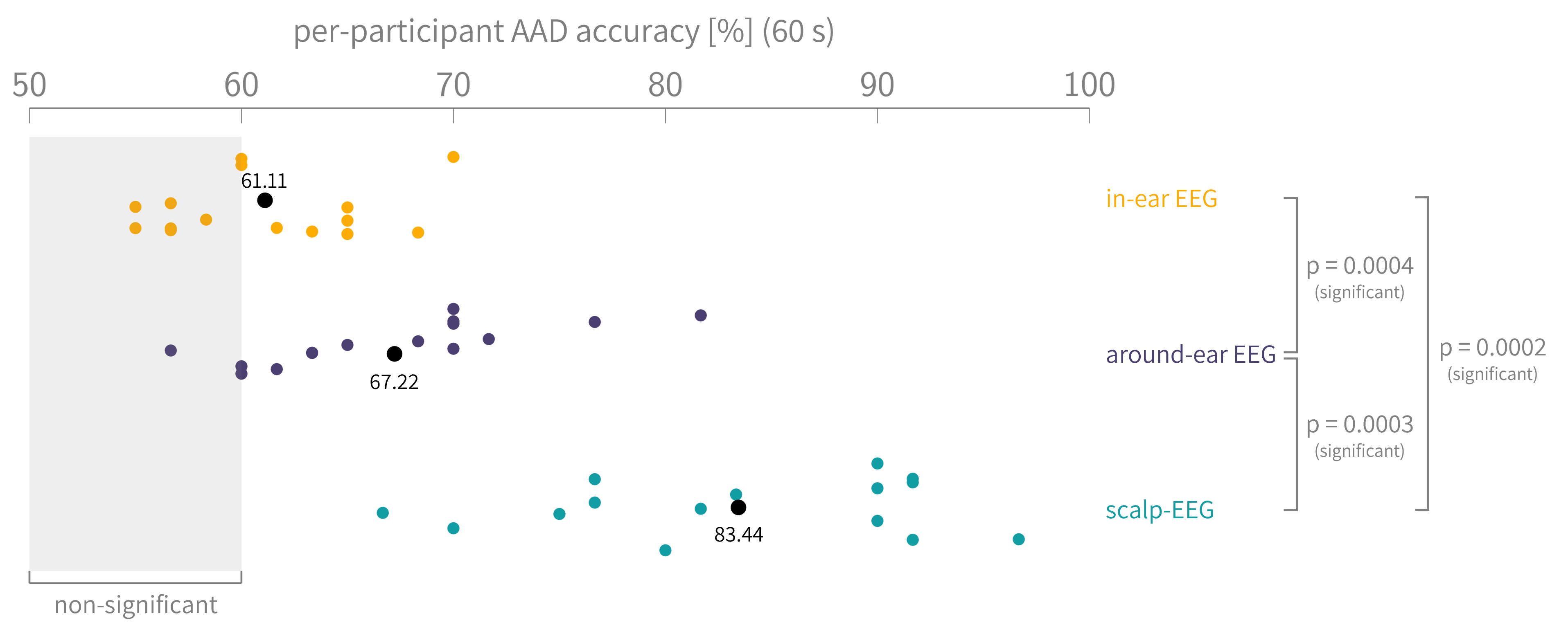}
			\caption{}
			\label{fig:per-subject-comparison}
		\end{subfigure}
		\caption{\textbf{(a)} The average performance curves (mean $\pm$ standard error on the mean; participant-specific decoding) show that all three setups achieve a significant average AAD accuracy across decision window lengths, while scalp EEG clearly outperforms the other ear-based setups. \textbf{(b)} The per-participant AAD accuracies using $\SI{60}{\second}$ decision windows show significant differences between all three setups.}
		\label{fig:baseline-performance}
	\end{figure}
	\subsection*{Comparison scalp, around-ear, and in-ear EEG}
	To assess the effectiveness of different EEG setups for AAD, we compare average and individual decoding performance across scalp, around-ear, and in-ear configurations. Figure~\ref{fig:baseline-comparison-car} shows the average performance curves (mean $\pm$ standard error on the mean) across participants for all three EEG setups with participant-specific decoding. The per-participant individual AAD accuracies on $\SI{60}{\second}$ decision windows are shown in Figure~\ref{fig:per-subject-comparison}, where the individual significance level corresponds to $60\%$ (higher than in Figure~\ref{fig:baseline-comparison-car} due to the smaller number of decisions for an individual participant). Clearly, all three setups can be used to significantly decode selective auditory attention to speech on average (Figure~\ref{fig:baseline-comparison-car}). On a per-participant basis, all participants yield significant results using scalp EEG, while this is the case for 14 out of 15 for around-ear EEG, and 9 out of 15 for in-ear EEG. However, using the Benjamini-Hochberg correction for multiple comparisons in in-ear EEG, only 6 out of 15 participants still yield a significant decoding accuracy. Similar multiple comparisons correction for scalp and around-ear EEG yields still significant accuracies for, respectively, (all) 15 and 13 participants.
	
	First, the results demonstrate that AAD can be effectively performed based on in-ear and around-ear EEG, confirming previous studies with similar accuracies for scalp and around-ear EEG~\cite{geirnaert2021eegBased,mirkovic2016target,holtze2022ear,fiedler2017single,thornton2024comparison}. Importantly, this shows that AAD is also possible using a fully in-ear EEG setup, even without external reference, achieving $61.11\%$ accuracy on $\SI{60}{\second}$ decision windows. At the same time, on an individual level, significant performance is only observed in six participants. This suggests that individual variability in, e.g., attentional performance, neural entrainment, earpiece fit, and artifacts can substantially affect decoding success.
	
\begin{figure}
		\centering
		\includegraphics[height=0.275\textheight]{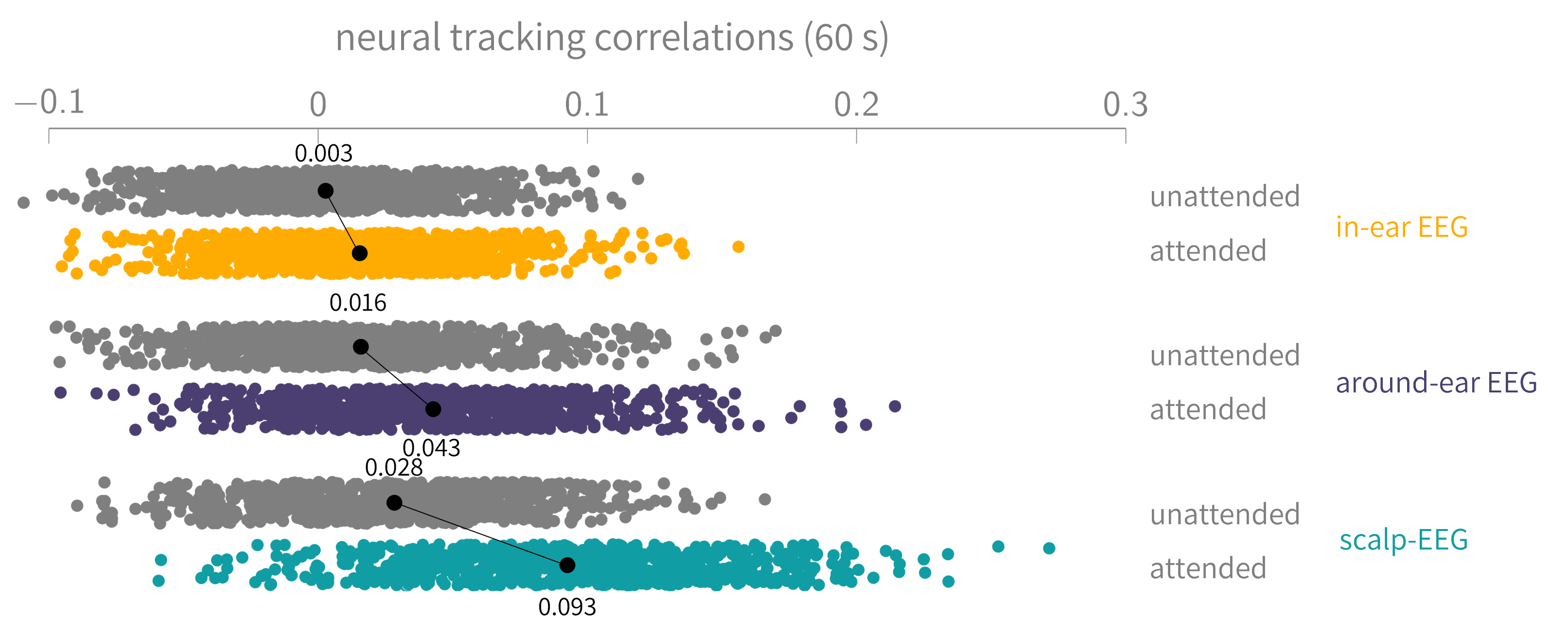}
		\caption{The neural tracking correlations (on $\SI{60}{\second}$ decision windows with participant-specific decoders) show enhanced neural tracking with both the attended and unattended speech when moving from in-ear to around-ear and scalp EEG, with a larger difference in average correlation between competing speakers, resulting in better decoding accuracies.}
		\label{fig:neural-tracking-correlations}
	\end{figure}

	Furthermore, clear differences in performance can be observed between all three EEG setups. Pairwise Wilcoxon signed rank tests ($n = 15, \alpha$-level $=0.05$, Benjamini-Hochberg multiple comparisons correction) reveal significant differences between all three setups for the $\SI{60}{\second}$-accuracies (Figure~\ref{fig:per-subject-comparison}). As expected, there is a clear trade-off between performance, driven by spatial coverage of the EEG electrodes, and wearability and unobtrusiveness of the EEG setups. The neural tracking correlations with attended and unattended speech, shown in Figure~\ref{fig:neural-tracking-correlations}, explain the observed performance gaps. When moving from in-ear to around-ear and scalp EEG, the difference between the average neural tracking correlations of the attended and unattended envelope increases, while the standard deviation remains more or less the same. Interestingly, both the attended and unattended envelope can be better decoded from spatially more diverse EEG setups, yet, this increase is much higher for the attended envelope. As shown in Lopez-Gordo et al.~\cite{gordo2025unsupervised}, when moving to longer decision windows, the average correlations remain constant, while the standard deviation further decreases. This explains why all setups in Figure~\ref{fig:baseline-comparison-car} converge towards $100\%$ when extending the decision window.
	
	To assess the impact of EEG preprocessing, especially for ear-based setups, we re-evaluate AAD accuracies on $\SI{60}{\second}$ decision windows (Figure~\ref{fig:per-subject-comparison}) after omitting most artifact removal steps, namely EOG regression (for scalp EEG), bad channel removal, ASR, and high-power artifact rejection. The average AAD accuracy increases to $86.11\%$ for scalp EEG (all 15 participants significant, with and without Benjamini–Hochberg correction), but decreases to $64.89\%$ for around-ear EEG (11 participants significant with and without correction) and to $58.78\%$ for in-ear EEG (with 4/7 significant accuracies with/without correction). The substantial performance drop in AAD accuracy and number of participants with significant performance for both around-ear and in-ear EEG demonstrates the higher importance of more rigorous artifact removal for achieving reliable decoding with ear-based EEG. All results reported in the remainder of this paper are therefore based on the full preprocessing pipeline.
	
	The neuro-steered hearing device application requires rapid decision-making to quickly adapt speaker gains to attention shifts. While Figure~\ref{fig:baseline-comparison-car} and the MESD (median $\SI{35.4}{\second}$ for scalp EEG) suggest that even scalp EEG - using linear stimulus reconstruction - might not perform accurately enough at very short decision window lengths, this is markedly worse for around-ear and especially in-ear EEG, with median MESDs of $\SI{2.66}{\minute}$ and $\SI{7.35}{\minute}$ respectively. However, Figure~\ref{fig:baseline-comparison-car} also shows that ear-based setups can be used in attention decoding use cases that are less speed-sensitive or allow offline processing.
	
	The reduced AAD performance observed in in-ear EEG may be attributed to several factors affecting the SNR. For clarity in the following discussion, we define “signal” as the electrophysiological response associated with auditory stimuli and the AAD task, and “noise” as everything else. First, the noise can stem from (at least) four fundamentally different sources: 
	\begin{enumerate}
		\item Background EEG unrelated to the AAD task. Both signal and this type of noise scale similarly with electrode distance, leaving the SNR approximately constant.
		\item Physiological artifacts, such as those caused by motion, muscle activity, eye or jaw movements, and sweating. Their influence depends on electrode placement and configurations. The use of individually customized earpieces made from flexible silicone helps maintain consistent electrode–skin contact pressure, reducing motion susceptibility. Still, in-ear EEG may be affected more by jaw movements, but, in contrast, less by others~\cite{kappel2017physiological}. Given the experiment was conducted under well-controlled conditions in which the participants sat still and refrained from speaking, we expect this contribution to be minor.
		\item Measurement noise, such as thermal noise and amplifier current noise. These scale with electrode impedance and are independent of electrode distances. Unlike the scalp and around-ear EEG setups, the in-ear electrodes are \emph{dry} and, therefore, have higher impedances. This leads to increased measurement noise, which will only impact the SNR substantially if the measurement noise is larger or close to the noise contributions from the other sources. The impact of thermal noise from the electrodes is discussed in detail in Kappel et al.~\cite{kappel2022characterization} and is unlikely to be a dominant contribution to the total noise in this study.
		\item Common-mode noise, which becomes differential noise due to finite common-mode rejection. The common-mode rejection depends on impedance imbalance, which increases with increasing electrode impedance. Consequently, dry-contact in-ear EEG is more prone to common-mode interference, likely contributing substantially to a reduced SNR. 
	\end{enumerate}
	Second, the signal part comprises projections of AAD-relevant neural sources from the brain to the ear. Kappel et al.~\cite{kappel2019ear} proposed a forward model that describes these projections from source to measurement space for in-ear EEG. They demonstrated that these projections to the ear heavily depend on both the distance and spatial orientation of the sources. The SNR is determined by the aggregated sum of projections from AAD-relevant neural sources relative to the aggregated sum of projections from all other brain sources. This ratio may be less favorable for in-ear EEG - when used in isolation without other external electrodes - compared to other electrode positions.
	
\begin{figure}
		\centering
		\includegraphics[height=0.35\textheight]{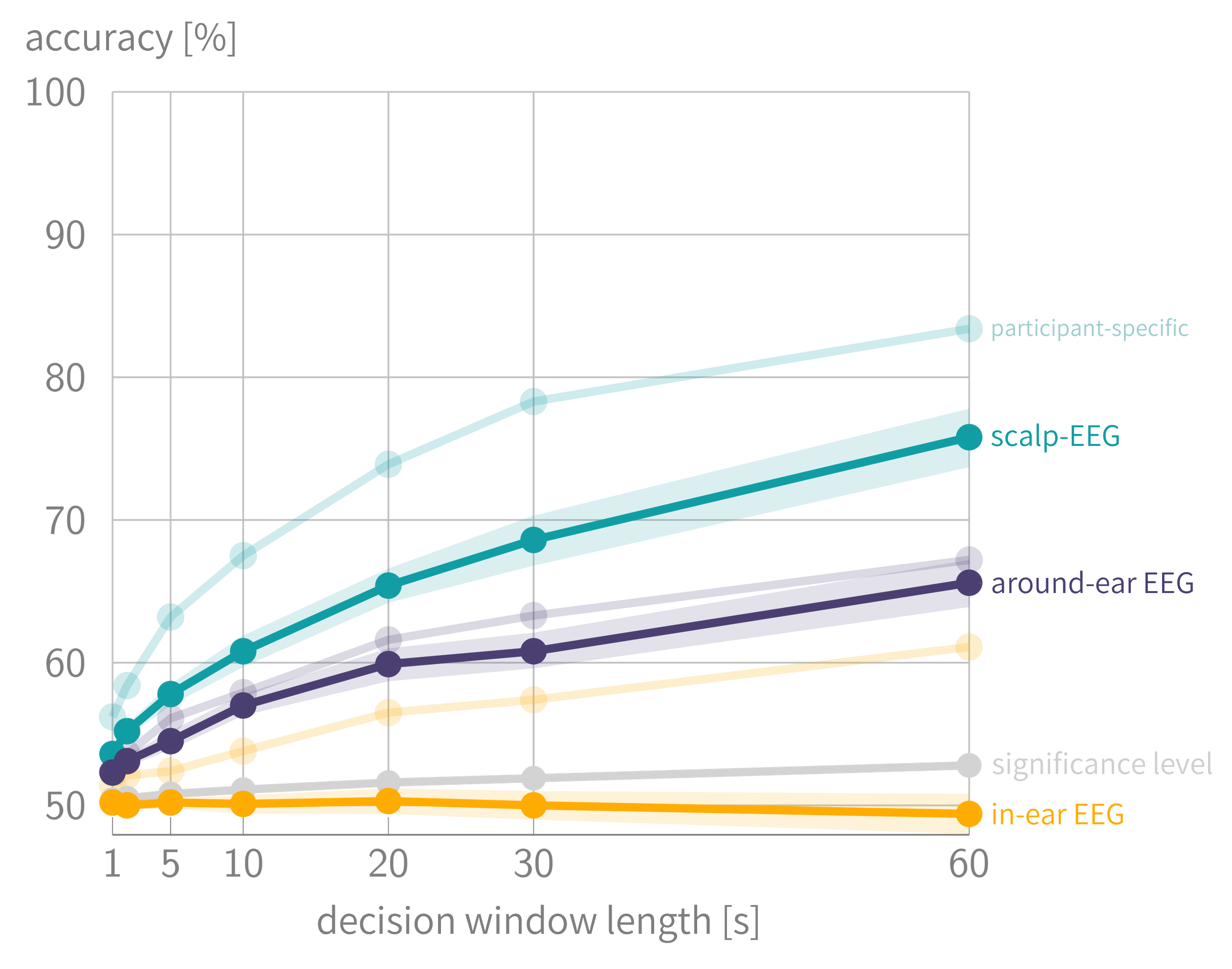}
		\caption{The average performance curves (mean $\pm$ standard error on the mean) of a participant-independent decoder show non-significant performance when using in-ear EEG, while generalizing across participants results in only a very small performance drop for around-ear EEG w.r.t. participant-specific decoders. The shaded performance curves show the performance curves from Figure~\ref{fig:baseline-comparison-car} with participant-specific decoders to facilitate comparison.}
		\label{fig:ptau-subject-independent}
	\end{figure}

	Figure~\ref{fig:ptau-subject-independent} shows the average performance curves (mean $\pm$ standard error on the mean) when using a participant-independent decoder, evaluated with leave-one-participant-out CV. Using scalp EEG, the typical decrease in accuracy w.r.t. participant-specific decoding is observed, with a mean $7.67$ percent point decrease on $\SI{60}{\second}$ windows, similarly to what was observed in Geirnaert et al.~\cite{geirnaert2021unsupervised}. For in-ear EEG, such generalization across participants was unsuccessful, as below-significance accuracies are obtained, with a $11.67\%$ percent point decrease w.r.t. participant-specific decoding. A possible explanation is that the training set in the participant-independent case for in-ear EEG contains much more `noise', contributed by those participants that did not obtain significant performance in the participant-specific case either. For example, using only the six participants that show significant AAD accuracy with in-ear EEG (after correction), a $57.78\%$ AAD accuracy on $\SI{60}{\second}$ windows could be achieved with in-ear EEG (3 out of 6 participants significant). Moreover, differences in ear anatomy result in different personalized earpieces, with sometimes different orientations for specific EEG electrodes. Lastly, given the previous discussion on the SNR of in-ear EEG, the amount and nature of the noise could vary more between participants. Surprisingly, the around-ear EEG setup results in a much smaller decrease in performance (mean $1.67$ percent point) for participant-independent decoding. We hypothesize this is due to the lower inter-participant variability in electrode positioning with the standardized C-shaped electrode array, compared to the highly personalized and variable in-ear setups, while the spatial diversity is also smaller than in scalp EEG, leading to less diverse EEG patterns.
	
	\subsection*{Complementarity of EEG setups}
	
\begin{figure}
		\centering
		\includegraphics[height=0.35\textheight]{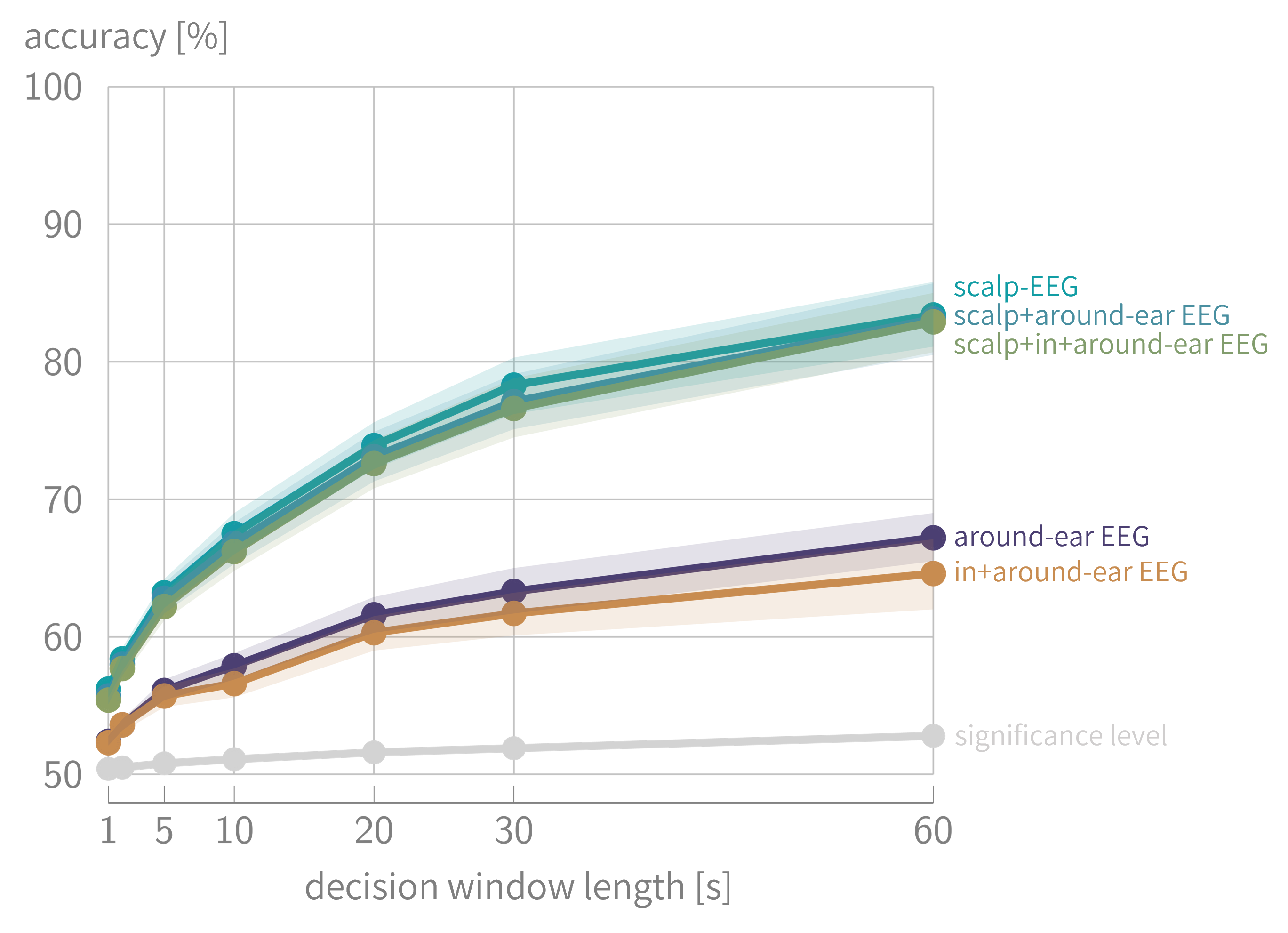}
		\caption{The average performance curves (mean $\pm$ standard error on the mean, participant-specific decoding) of the various combined EEG setups show no added benefit from incorporating more wearable setups.}
		\label{fig:complementarity}
	\end{figure}

	Figure~\ref{fig:complementarity} shows the average performance curves (mean $\pm$ standard error on the mean, participant-specific decoding) when combining more wearable and unobtrusive EEG setups with a bulkier, spatially more diverse setup to explore the complementarity of the various EEG setups. When combining scalp EEG with around-ear and/or in-ear EEG, all setups are first re-referenced to the shared Fp1 electrode to bring them in the same `reference' system and remove amplifier-specific effects. Early integration is used, meaning that the EEG channels are combined before the decoder is trained. It is important to note that preprocessing was not re-executed when combining the setups, which could, for example, improve ASR.
	
	The results show no benefit of adding wearable setups, as no improvements in performance are present. The stimulus decoder is thus not able to improve decoding by leveraging additional channels from around or in the ear. This lack of improvement is expected, given the significant performance gaps observed in Figure~\ref{fig:baseline-performance} and the neural tracking correlations in Figure~\ref{fig:neural-tracking-correlations}. From the perspective of complementing bulkier setups with wearable ear-setups, no AAD performance gains are thus attained.
	
	\subsection*{Influence of the reference}
	
\begin{figure}
		\centering
		\begin{subfigure}{0.5\linewidth}
			\centering
			\includegraphics[width=1\linewidth]{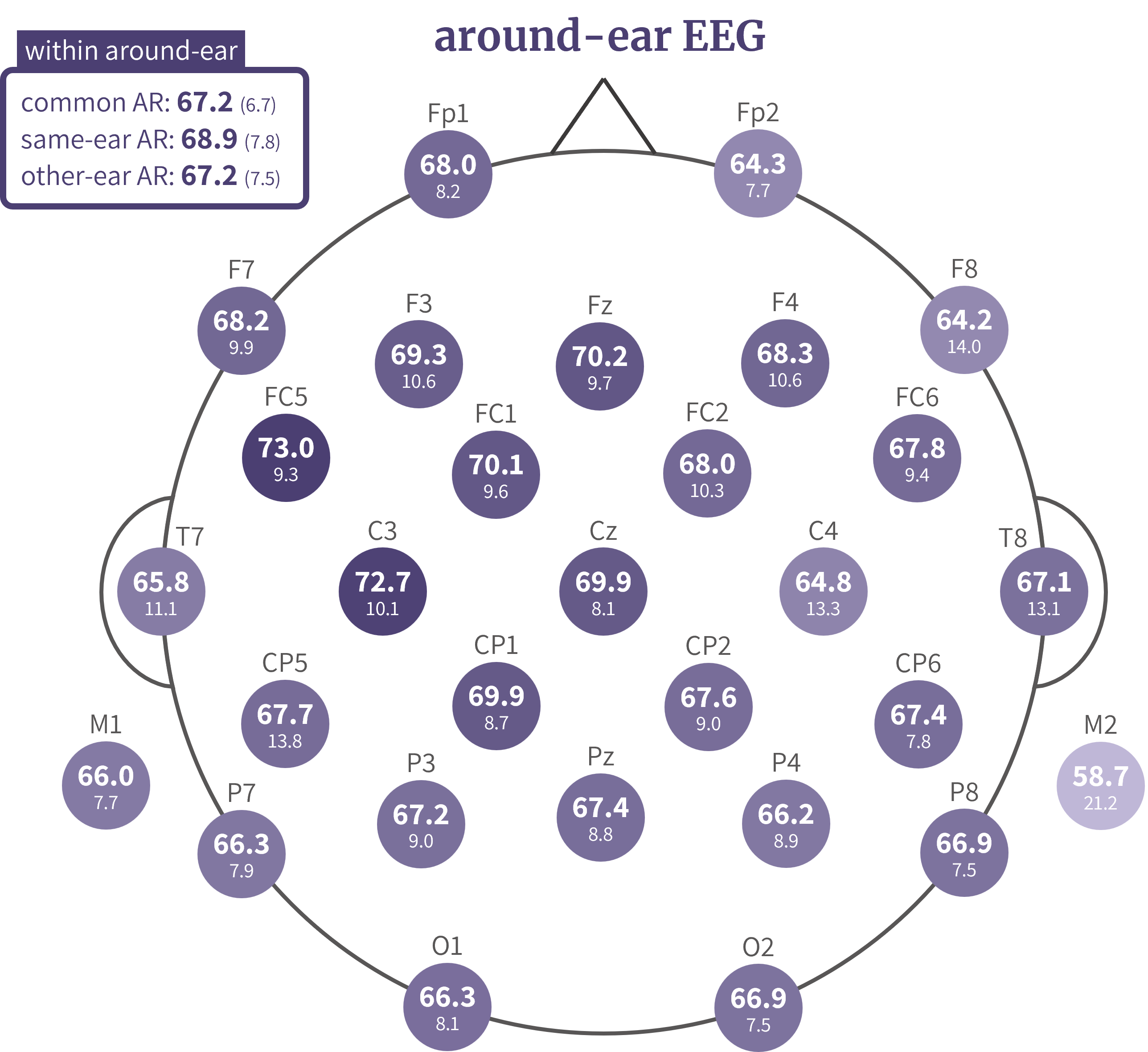}
			\caption{}
			\label{fig:reference-around-ear}
		\end{subfigure}%
		\begin{subfigure}{0.5\linewidth}
			\centering
			\includegraphics[width=1\linewidth]{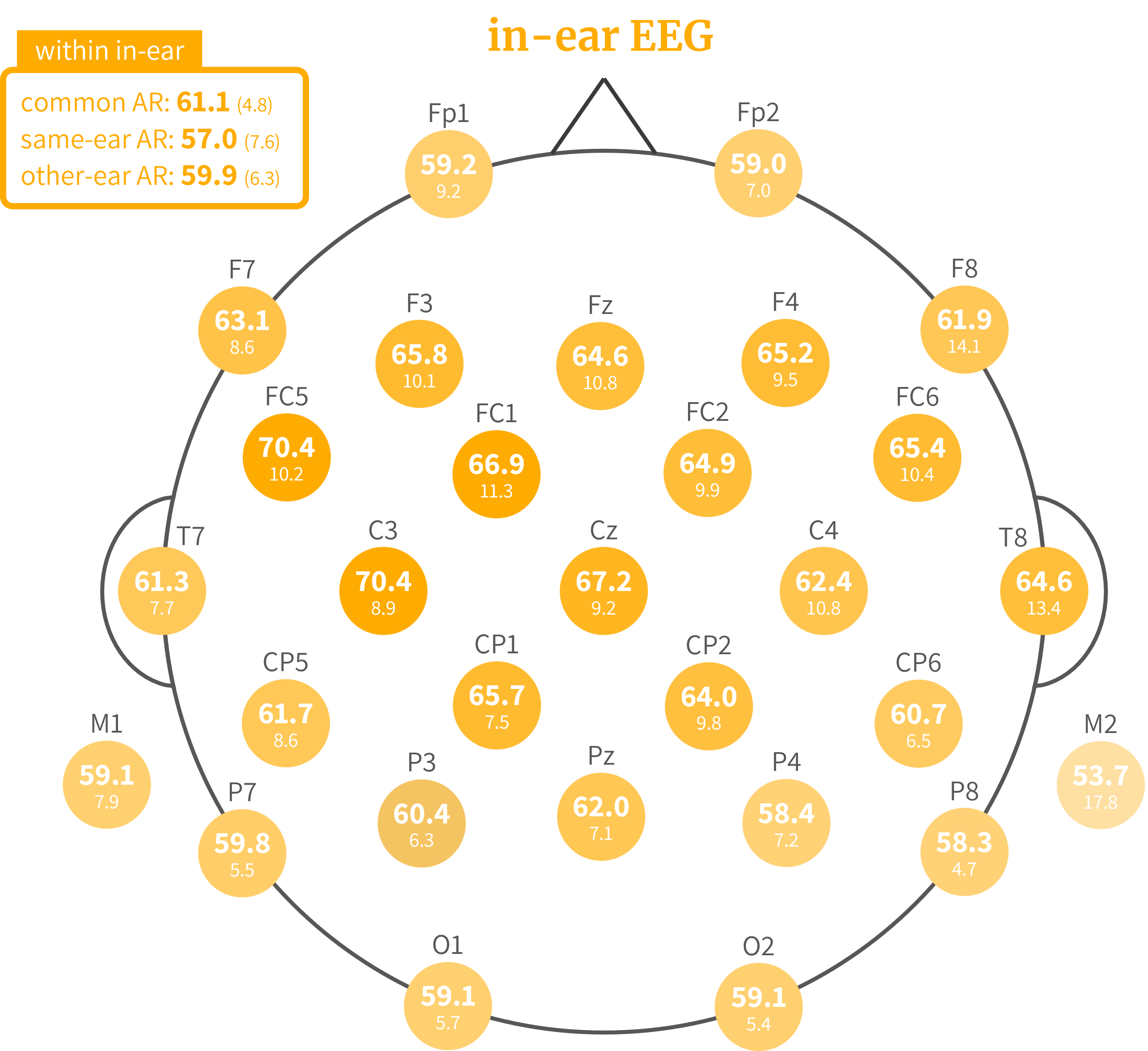}
			\caption{}
			\label{fig:reference-in-ear}
		\end{subfigure}
		\caption{The mean AAD accuracy (participant-specific decoding) on $\SI{60}{\second}$ decision windows (with standard deviation indicated below) when using common average referencing (AR), same-ear AR, and other-ear AR, and when re-referencing to individual scalp electrodes, for \textbf{(a)} around-ear EEG and \textbf{(b)} in-ear EEG. The best external scalp reference electrodes are found in the left-lateralized fronto-central area.}
		\label{fig:reference}
	\end{figure}

	Figure~\ref{fig:reference} shows the influence of re-referencing on the AAD accuracy for $\SI{60}{\second}$ decision windows (with participant-specific decoding), comparing alternative referencing within the ear-based systems (same-ear and other-ear average reference (AR)) and external scalp reference electrodes. Same-ear AR corresponds to re-referencing to the average channel within the same ear, and other-ear AR to re-referencing to the average channel of the other ear. The box shows alternative \emph{within}-ear referencing, while the scalp heatmaps show AAD accuracy when referencing to each individual scalp electrode, with standard deviation indicated below.

	Within-system re-referencing allow the ear-based EEG systems to remain truly `ear'-based systems in the strict sense. Especially interesting is same-ear AR, as it allows each ear to be galvanically isolated, removing the practical requirement for wired connections between ears, thereby, improving unobtrusiveness and susceptibility to artifacts. Such an approach was, for example, taken for ASSR detection in Ding et al.~\cite{ding2025synchronized}. For in-ear EEG, this leads to a performance drop of $4.1$ percent point, whereas around-ear EEG shows a small increase ($1.7$ points). This suggests that in-ear EEG is more sensitive to this galvanic isolation, while around-ear EEG performs comparably well with local referencing.
	
	Using the external scalp electrodes as references, we can now compare our results to the prior studies of Fiedler et al.~\cite{fiedler2017single} and Thornton et al.~\cite{thornton2024comparison}, both using FT7 as an external reference. Fiedler et al.~\cite{fiedler2017single} reported $70\%$ accuracy using $\SI{60}{\second}$ decision windows on 4 participants, while Thornton et al.~\cite{thornton2024comparison} reported around $63.5\%$ using $\SI{30}{\second}$ decision windows on 18 participants. While we do not have the FT7 electrode available in our setup, with FC5, close to FT7, we obtain comparable accuracies of $70.4\%$ ($\SI{60}{\second}$) and $64.9\%$ ($\SI{30}{\second}$), confirming the results found in these respective papers. Note, however, that in both cited studies, only one in-ear electrode per ear (with only the left ear used in Fiedler et al.~\cite{fiedler2017single}) is used, while here, we have six in-ear electrodes per ear. This suggests limited benefit from the increased spatial diversity within the ear when using a longer-distance, external scalp reference electrode. Note that when shorter-distance, within ear-setup referencing is used, a benefit of an increased number of in ear-electrodes is expected, as, for example, shown in ASSR estimation~\cite{sergeeva2022investigation}.
	
	The heatmaps in Figure~\ref{fig:reference} furthermore show that the most effective scalp references are located in the left-lateralized fronto-central areas, with highest performances for the electrodes FC5, C3, FC1, and Cz. A similar finding was established in Fiedler et al.~\cite{fiedler2017single}. One explanation is that a reference in this area creates an axis through the auditory cortex, which has shown to be important (as physiologically expected) in stimulus reconstruction using data-driven channel selection on scalp EEG~\cite{mirkovic2015decoding,narayanan2020analysis,narayanan2021eeg,nguyen2024aadnet}. In general, adding a single scalp reference electrode to the in-ear setup largely improves AAD accuracy by $9.3$ percent point w.r.t. within in-ear referencing. Such a strategically placed additional electrode thus offers interesting opportunities to improve decoding, a strategy that is further explored in the next section. For the around-ear setup, this impact is smaller ($5.8$ percent point), yet substantial.
	
	\subsection*{Improving ear-based EEG for AAD with scalp electrodes}
	The previous section highlighted the potential of enhancing AAD performance by adding a single scalp reference electrode. Here, we further explore how combining scalp electrodes with ear-based EEG setups could improve performance. We adopt the perspective of an EEG sensor network, where different sensor nodes, representing miniaturized EEG devices, can be flexibly combined~\cite{narayanan2020analysis,narayanan2021eeg}. These nodes can include in-ear and around-ear EEG, as well as other flex-printed sensors or tattoo electrodes~\cite{casson2019wearable}. In this analysis, the in-ear and around-ear configurations are treated as distinct nodes, while each scalp electrode is also considered an individual node. All signals are first re-referenced to the shared Fp1 electrode to enable consistent integration across nodes (which are therefore not galvanically isolated as in a wireless EEG sensor network). As a result, the Fp1 electrode is excluded from node selection.
	
	To investigate how AAD accuracy evolves when nodes are added to either the in-ear or around-ear setup, we employ a greedy forward node selection strategy. At each step, the node that maximally improves AAD accuracy on a validation set is added. The selection process also allows the other ear-based setup as a candidate node. This forward selection, using participant-specific decoding, is conducted using an inner leave-one-trial-out CV loop nested within the outer leave-one-trial-out CV used for evaluation. Node selection is based on AAD accuracy over $\SI{30}{\second}$ decision windows on the left-out validation trials in the inner CV procedure, which is shorter than the $\SI{60}{\second}$ windows used during testing to reduce the number of ties. Note that this greedy strategy due to its sequential nature has a limited ability to consider interactions between multiple node combinations at once, which could potentially lead to higher performance gains. As a control condition to assess the validity of the forward node selection, a random node selection procedure is implemented. For each participant and leave-one-trial-out CV fold, 100 repetitions of sequentially adding randomly selected nodes are performed.
	
\begin{figure}
		\centering
		\includegraphics[height=0.35\textheight]{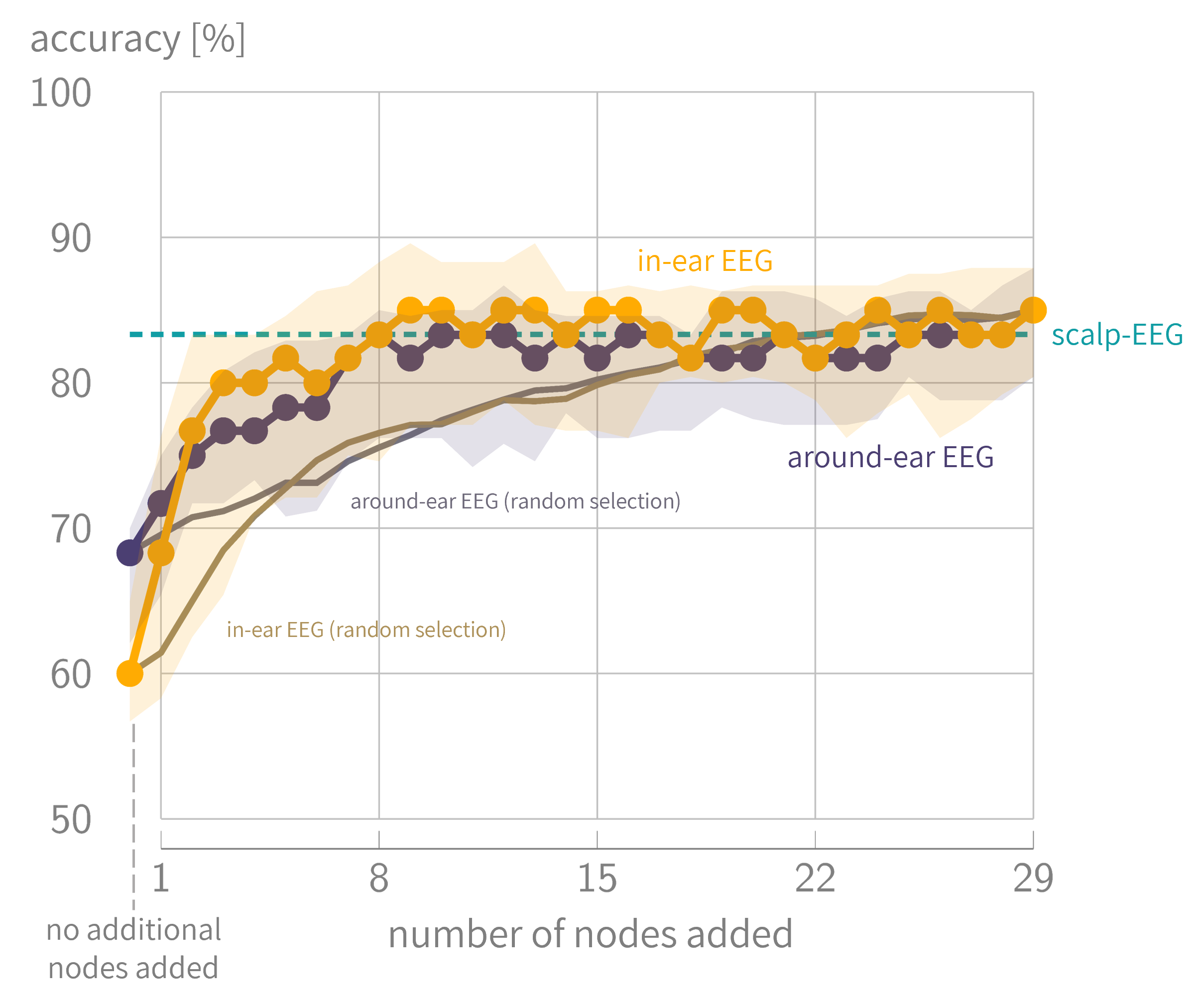}
		\caption{The average accuracy (median with $25\%$- and $75\%$-percentiles, participant-specific decoding) on $\SI{60}{\second}$ decision windows as nodes are sequentially added to in- and around-ear EEG in a greedy foward selection shows convergence to scalp EEG performance after adding 8 nodes. The average accuracy (median across participants, mean across 100 random repetitions) when using random node selection is added as a baseline.}
		\label{fig:nodeSelection-performance}
	\end{figure}

	Figure~\ref{fig:nodeSelection-performance} show the average accuracy across participants (median with $25\%$- and $75\%$-percentiles, participant-specific decoding) on $\SI{60}{\second}$ decision windows when sequentially adding more nodes, starting from the original in-ear and around-ear setups. The average accuracy (median across participants, mean across repetitions) obtained from the random node selection is also shown as a control. This comparison supports the validity of the greedy forward node selection strategy, which consistently outperforms the random baseline, particularly (and as expected) when fewer nodes are added. For both ear-based setups, performance reaches full-scalp EEG performance after adding eight additional nodes, with diminishing returns beyond each added node. Notably, the in-ear setup achieves near-scalp median accuracy with only three additional nodes, while the around-ear setup shows a more gradual and less steep improvement. Between two and seven added nodes, the in-ear sensor network consistently outperforms the around-ear configuration. This observation aligns with the reference electrode analysis, which also suggested that external electrodes have a greater impact on the in-ear setup. One explanation is that the around-ear configuration already performs relatively well, but another lies in spatial diversity: the in-ear setup offers more distinct spatial information, being positioned further from conventional scalp electrodes. While the complementarity analysis showed no benefit of in-ear (or around-ear) EEG added to full-scalp EEG, these findings suggest that in-ear EEG is highly suitable as a node in compact EEG sensor networks based on a few scalp electrodes.
	
\begin{figure}
		\centering
		\begin{subfigure}{0.5\linewidth}
			\centering
			\includegraphics[width=1\linewidth]{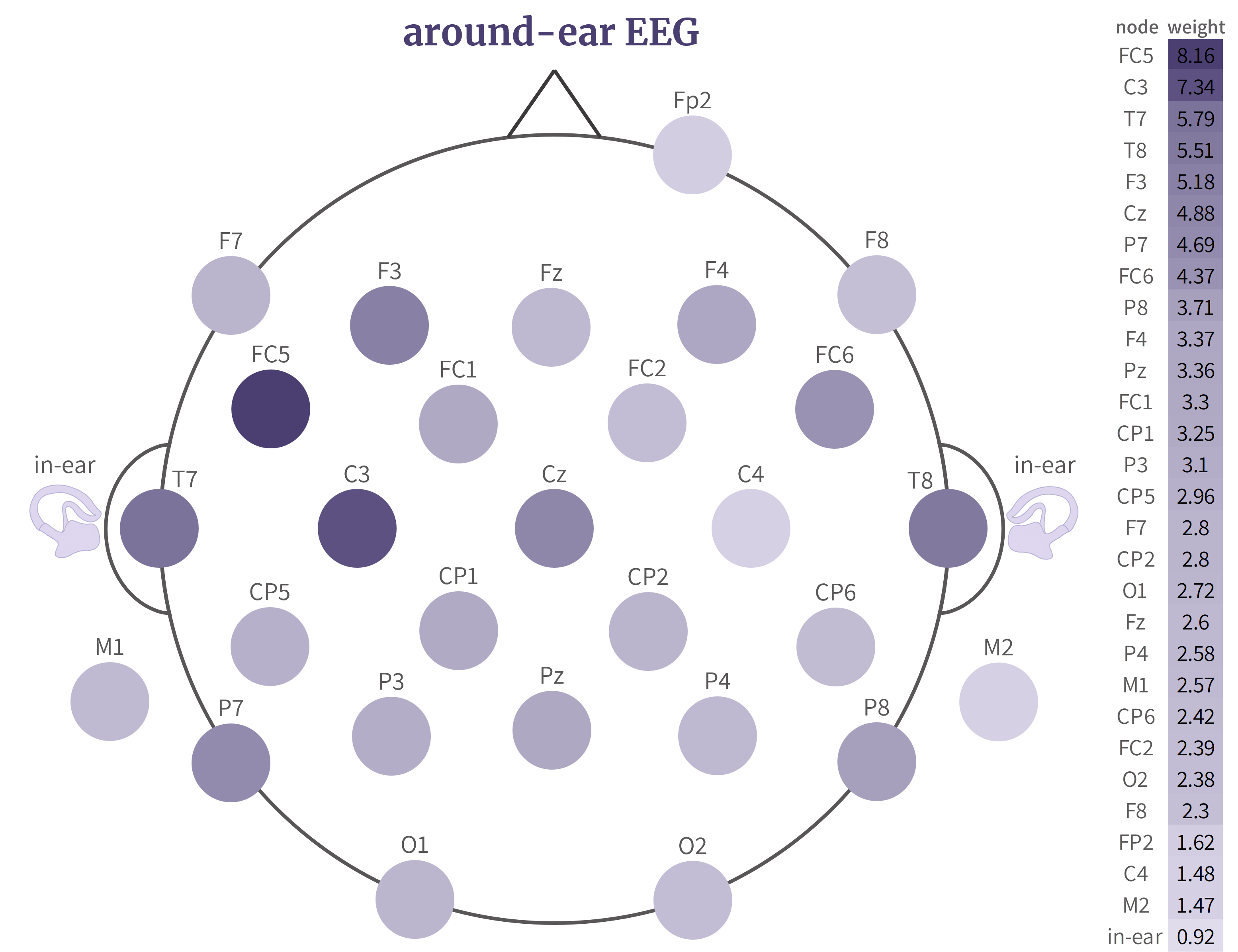}
			\caption{}
			\label{fig:importance-nodeSelection-aroundear}
		\end{subfigure}%
		\begin{subfigure}{0.5\linewidth}
			\centering
			\includegraphics[width=1\linewidth]{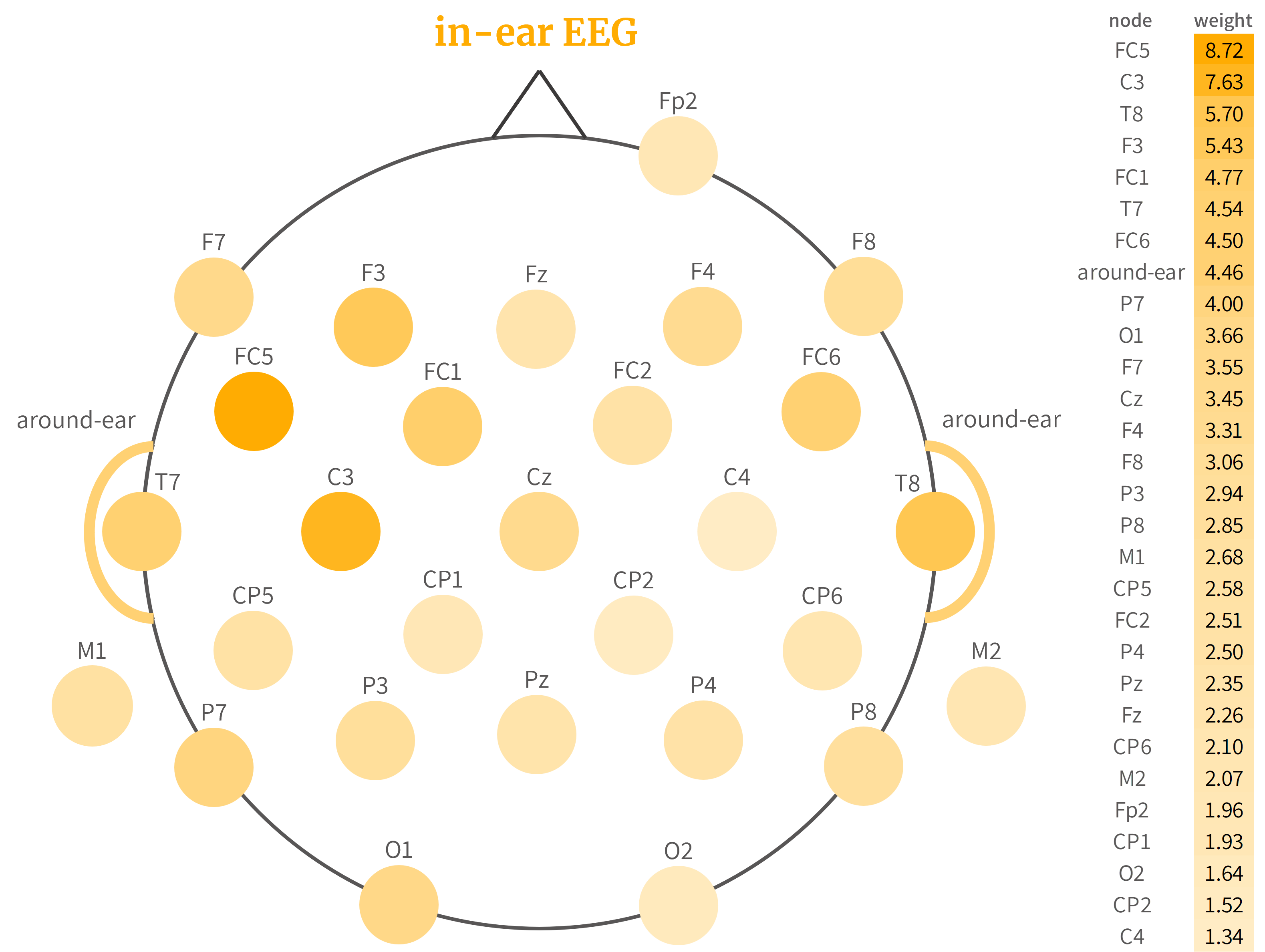}
			\caption{}
			\label{fig:importance-nodeSelection-inear}
		\end{subfigure}
		\caption{The heatmaps and weights of selected additional nodes starting from \textbf{(a)} around-ear and \textbf{(b)} in-ear EEG show that the most popular additional nodes are positioned in the fronto-central and temporal brain areas. Weights are determined per fold and participant using exponential decay (base 2, half-life 3), normalized to sum to $100$, and averaged across folds and participants.}	
		\label{fig:importance-nodeSelection}
	\end{figure}

	Figure~\ref{fig:importance-nodeSelection} provides a heatmap of selected nodes, based on their importance in the greedy selection. For each fold and participant, nodes are assigned a weight according to their selection order, using an exponential decay with a base of 2 and a half-life of three nodes (i.e., the third-selected node receives half the weight of the first). This half-time is inspired by the diminishing returns in the in-ear accuracy curve in Figure~\ref{fig:nodeSelection-performance}. The weights are scaled to sum to $100$ and are then averaged across folds and participants. For both around-ear and in-ear EEG, the most `popular' additional nodes are located in the fronto-central and temporal areas. Specifically, FC5, C3, and T7 are among the three most popular nodes for the around-ear setup (Figure~\ref{fig:importance-nodeSelection-aroundear}), while FC5, C3, and T8 are top picks for in-ear EEG (Figure~\ref{fig:importance-nodeSelection-inear}) — consistent with the earlier findings of the reference analysis in Figure~\ref{fig:reference}. Interestingly, when starting from the in-ear setup, the around-ear node is the 8\textsuperscript{th} most popular additional node, whereas, as expected, the in-ear node is the least popular node when starting from around-ear EEG.
	
	To ensure that the in-ear EEG node contributes meaningfully to the sensor network when only a few scalp electrodes are selected, we compare performance with and without its inclusion. Specifically, combining the in-ear node with the top three scalp nodes selected for this configuration (FC5, C3, T8; see Figure~\ref{fig:importance-nodeSelection-inear}) yields a mean AAD accuracy of $73.4\%$ on $\SI{60}{\second}$ windows. In contrast, the same three scalp electrodes without the in-ear node achieve only $64.0\%$. A Wilcoxon signed-rank test ($n = 15$, $\alpha$-level $= 0.05$) confirms a significant difference between both configurations ($p = 0.0123$), validating the added value of in-ear EEG node within a minimal EEG sensor network for AAD.

	\section*{Conclusion}
	In this paper, we presented and analyzed a novel auditory attention decoding (AAD) dataset in which scalp, around-ear, and dry-contact in-ear EEG were recorded simultaneously for the first time. Notably, this is also the first AAD analysis using exclusively in-ear EEG, without the inclusion of an external (scalp) reference electrode.
	
	Using a classic linear stimulus reconstruction algorithm, we demonstrated significant AAD performance gaps between the three setups, with average AAD accuracies of $83.44\%$, $67.22\%$, and $61.11\%$ on $\SI{60}{\second}$ decision windows for scalp, around-ear, and in-ear EEG, respectively. Both around-ear and especially in-ear EEG showed reduced decoding performance (correlation) for attended and unattended speech features. These performance gaps underscore the clear trade-off between decoding accuracy and the wearability, concealability, and unobtrusiveness of the EEG setup. Importantly, scalp and around-ear EEG used wet electrodes, reducing their practical usability, while in-ear EEG used dry-contact electrodes. Due to the large performance gaps, ear-based EEG did not enhance performance when combined with full-scalp EEG. Additionally, participant-independent decoding proved particularly challenging with in-ear EEG.
	
	While even scalp EEG with linear stimulus reconstruction may not yield sufficient accuracy at short decision window lengths for decision speed-sensitive applications like AAD for neuro-steered hearing devices, our results show that this limitation is, with the current preprocessing and decoding approach, even more pronounced for ear-based EEG, and particularly in-ear EEG. Nonetheless, we showed that selective auditory attention can still be decoded from both around-ear and in-ear EEG when longer decision windows are used, indicating the potential of these highly concealable and unobtrusive systems for attention monitoring over longer timescales.
	
	Importantly, we found that adding a scalp reference electrode or a small number of scalp electrodes in a greedy forward selection leads to rapid improvements in AAD accuracy. When starting from in-ear EEG, near-scalp performance was achieved by adding just three scalp electrodes, and full scalp-level accuracy was reached by adding eight electrodes to either in-ear or around-ear setups. Interestingly, in-ear EEG benefited more and faster than around-ear EEG from the addition of the first few scalp electrodes, highlighting its potential value as part of an EEG sensor network, complemented with a few other strategically placed EEG sensor nodes.
	
	While the current dataset reveals clear trends when comparing the different EEG configurations, the sample size of 15 participants remains relatively limited. Additionally, given the focus on the direct comparison of scalp-, around-ear, and in-ear EEG, a homogeneous group of young, normal-hearing participants was recruited. While literature has shown that linear stimulus reconstruction for AAD using scalp EEG is feasible in elderly people or individuals with hearing loss~\cite{fuglsang2020effects}, future studies should consider more heterogeneous participant groups to ensure generalizability to these populations. Moreover, to assess the full potential of wearable EEG systems, the experiment could be repeated in more ecologically valid listening scenarios. In that context, the dry in-ear EEG setup used here should be compared with other dry electrode configurations. Another limitation is that only one comprehension question was asked after each $\SI{10}{\minute}$-trial. This might not sufficiently motivate participants to sustain focus throughout the trial. Since task performance was probed only at the end of each trial and responses were not logged, fluctuations in participants’ attention could not be accounted for. Lastly, we employed a linear stimulus reconstruction AAD algorithm. While several nonlinear (e.g., deep learning-based) AAD methods have been developed for scalp EEG, such approaches have been lacking or unsuccessful for ear-based EEG~\cite{thornton2024comparison}. This highlights a clear need for the development of tailored nonlinear AAD algorithms for ear-based EEG.
	
	Finally, to support and facilitate further research and development, we have made this dataset publicly available. Future directions to enhance ear-based AAD performance include the development of more tailored preprocessing techniques and the application of transfer learning strategies to leverage scalp EEG data for improving ear-based decoding.
	
	\section*{Data availability}
	The EEG dataset generated and analyzed during the current study is publicly available in the following Zenodo repository~\cite{geirnaert2025ear}: \url{https://doi.org/10.5281/zenodo.16536441}.
	
	\bibliography{bib-v2}
	
	\section*{Author contributions statement}
	
	S.G. and P.K. conceived and designed the experiment(s) and study; S.G., S.L.K., and P.K. implemented the experiment; S.G. conducted the experiments, analyzed the results, and wrote the first manuscript draft. All authors reviewed and edited the manuscript. 
	
	\section*{Additional information}
	\subsection*{Acknowledgements}
	
	This research was supported by a junior postdoctoral fellowship fundamental research (for S. Geirnaert, grant No. 1242524N) and a travel grant for a long stay abroad (for S. Geirnaert, grant No. V418923N) from the Research Foundation Flanders (FWO). Equipment and laboratory facilities were provided by the Center for Ear-EEG at Aarhus University.
	
	S. Geirnaert is also affiliated to Leuven.AI - KU Leuven institute for AI, B-3000, Leuven, Belgium.
	
	The authors thank Nhan Duc Thanh Nguyen for providing help in implementing the protocol and Anna Sergeeva for providing help with the modeling and production of the earpieces.
	
	\subsection*{Competing interests}
	The authors declare no competing interests.

\end{document}